\documentclass[sigconf]{acmart}
\def\BibTeX{{\rm B\kern-.05em{\sc i\kern-.025em b}\kern-.08emT\kern-.1667em\lower.7ex\hbox{E}\kern-.125emX}}
    
\copyrightyear{2019} 
\acmYear{2019} 
\acmConference[CIKM '19]{The 28th ACM International Conference on Information and Knowledge Management}{November 3--7, 2019}{Beijing, China}
\acmBooktitle{The 28th ACM International Conference on Information and Knowledge Management (CIKM '19), November 3--7, 2019, Beijing, China}
\acmPrice{15.00}
\acmDOI{10.1145/3357384.3357902}
\acmISBN{978-1-4503-6976-3/19/11}

\usepackage[colorinlistoftodos]{todonotes}
\newcommand{\nop}[1]{}

\usepackage{booktabs}
\usepackage{xspace} 
\usepackage{multirow}
\usepackage{diagbox}
\usepackage{subfigure}
\usepackage{braket}
\usepackage[perpage]{footmisc}
\usepackage[skip=2pt]{caption}
\usepackage{atbegshi}
\AtBeginDocument{\AtBeginShipoutNext{\AtBeginShipoutDiscard}}

\graphicspath{{./img/}}

\newcommand{\dataSet}{\textsl}
\newcommand{\arterial}{\dataSet{$Arterial_{1\times3}$}\xspace}
\newcommand{\gridthree}{\dataSet{$Grid_{3\times3}$}\xspace}
\newcommand{\gridsix}{\dataSet{$Grid_{6\times6}$}\xspace}

\newcommand{\newYork}{\dataSet{$D_{New York}$}\xspace}
\newcommand{\jinan}{\dataSet{$D_{Jinan}$}\xspace}
\newcommand{\hangzhou}{\dataSet{$D_{Hangzhou}$}\xspace}

\newcommand{\methodFont}{\textsl}
\newcommand{\formula}{\methodFont{Fixedtime}\xspace}

\newcommand{\maxPressure}{\methodFont{MaxPressure}\xspace}
\newcommand{\individualRL}{\methodFont{Individual RL}\xspace}
\newcommand{\oneModel}{\methodFont{OneModel}\xspace}
\newcommand{\GCN}{\methodFont{GCN}\xspace}
\newcommand{\GAT}{\methodFont{CoLight}\xspace}
\newcommand{\GATEuclidean}{\methodFont{CoLight}\xspace}
\newcommand{\GATNode}{\methodFont{CoLight\text{-}node}\xspace}
\newcommand{\nips}{\methodFont{CGRL}\xspace}
\newcommand{\oneModelNeighbor}{\methodFont{Neighbor RL}\xspace}

\newcommand{\problemFont}{\mathit}
\newcommand{\problemSetFont}{\mathcal}
\newcommand{\states}{\problemSetFont{S}}
\renewcommand{\state}{\problemFont{s}}
\newcommand{\observations}{\problemSetFont{O}}
\newcommand{\observation}{\problemFont{o}}
\newcommand{\actions}{\problemSetFont{A}}

\newcommand{\actionVec}{\problemFont{\textbf{a}}}
\newcommand{\reward}{\problemFont{r}}
\newcommand{\policy}{\problemFont{\pi}}
\newcommand{\transition}{\problemSetFont{P}}
\newcommand{\decay}{\problemFont{\gamma}}

\begin{document}
\title{CoLight: Learning Network-level Cooperation \\ for Traffic Signal Control}
\author{Hua Wei$^{\dagger*}$, Nan Xu$^{\ddagger*}$, Huichu Zhang$^\ddagger$, Guanjie Zheng$^\dagger$, Xinshi Zang$^\ddagger$, Chacha Chen$^\ddagger$,}\thanks{*Equal contribution}
\par
\author{ Weinan Zhang$^\ddagger$, Yanmin Zhu$^\ddagger$, Kai Xu$^\S$, Zhenhui Li$^\dagger$}
\affiliation{$^\dagger$Pennsylvania State University, $^\ddagger$Shanghai Jiao Tong Univerisity, $^\S$Shanghai Tianrang Intelligent Technology Co., Ltd\\
$^\dagger$\{hzw77, gjz5038, jessieli\}@ist.psu.edu,
$^\ddagger$\{xunannancy, chacha1997, wnzhang, yzhu\}@sjtu.edu.cn, $^\ddagger$\{zhc,xszang\}@apex.sjtu.edu.cn, $^\S$kai.xu@tianrang-inc.com}

\begin{abstract}
Cooperation among the traffic signals enables vehicles to move through intersections more quickly. Conventional transportation approaches implement cooperation by pre-calculating the offsets between two intersections. Such pre-calculated offsets are not suitable for dynamic traffic environments.

To enable cooperation of traffic signals, in this paper, we propose a model, \textit{CoLight}, which uses graph attentional networks to facilitate communication. Specifically, for a target intersection in a network, \textit{CoLight} can not only incorporate the temporal and spatial influences of neighboring intersections to the target intersection, but also build up index-free modeling of neighboring intersections. To the best of our knowledge, we are the first to use graph attentional networks in the setting of reinforcement learning for traffic signal control and to conduct experiments on the large-scale road network with hundreds of traffic signals.
In experiments, we demonstrate that by learning the communication, the proposed model can achieve superior performance against the state-of-the-art methods. 
\end{abstract}



\maketitle
{\fontsize{8pt}{8pt} 
\selectfont
\textbf{ACM Reference Format:}\\
Hua Wei, Nan Xu, Huichu Zhang, Guanjie Zheng, Xinshi Zang, Chacha Chen, and Weinan Zhang, Yanmin Zhu, Kai Xu, Zhenhui Li. 2019. CoLight: Learning Network-level Cooperation for Traffic Signal Control. In \textit{The 28th ACM International Conference on Information and Knowledge Management (CIKM '19), November 3--7, 2019, Beijing, China. } ACM, New York, NY, USA, 10 pages.
https://doi.org/10.1145/3357384.3357902.}
\section{Introduction}
A key question often asked of traffic signal control is ``How do traffic signals cooperate between intersections?'' Cooperation among intersections is especially important for an urban road network because the actions of signals could affect each other, especially when the intersections are closely connected. Good cooperation among the traffic signals enables the vehicles to move through intersections more quickly. 

In the transportation field, a typical way to solve the cooperation problem between intersections is to formulate it as an optimization problem and solve it under certain assumptions (e.g., uniform arrival rate~\cite{webster1958traffic, roess2004traffic} and unlimited lane capacity~\cite{Maxpressure}). Such methods, however, do not perform well because the assumptions do not hold in the real world.

Recently, researchers start to investigate reinforcement learning (RL) techniques for the cooperation of traffic signals. The most common way is to have an RL agent control each intersection and communication is achieved by sharing information among agents~\cite{EAA13,DrSt06,SBPB+06,ALUK10}. At each time step, the agent observes the traffic condition of the target intersection and its neighboring intersections, and decides an action $a$ to take. After the action is taken, a reward $r$ (often defined as a measure correlated with travel time) is fed back to the agent indicating how good the action $a$ is. Different from conventional approaches, such RL methods avoid making strong assumptions and directly learn good strategies from trials and errors.

Existing RL-based methods still fail to communicate with neighbors in the most efficient way. We propose \textit{CoLight} that improves communication of agents and is scalable to hundreds of intersections. In particular, our work makes the following key contributions:

\nop{Recent RL-based traffic signal control methods have shown to be more effective than conventional optimization-based transportation approaches. This paper extends this line of work by making several important new contributions:}

$\bullet$ 
\emph{Cooperation through dynamic communication}. Most methods to achieve cooporation are through expanding the observation of the target agent for more comprehensive information. Existing studies~\cite{EAA13,DrSt06,SBPB+06,ALUK10,chu2019multi} tend to select the traffic conditions from adjacent intersections and directly concatenate them with the traffic condition of the target intersection, neglecting the fact that the traffic is changing both temporally and spatially. For example, if intersections $A$ and $B$ are adjacent intersections on a major road, but intersection $C$ is on a side road linked with $B$. The information from $A$ is more useful to $B$ compared with that from $C$ to $B$. Besides, the influences from intersection $A$ could change temporally. For example, during the morning rush hour, there are a large number of vehicles moving from $A$ to $B$, and the direction of vehicles is reversed at night peak hours. Therefore, the effect of $A$ on $B$ is also changing at different times of the day. In this paper, we propose to use the graph attentional network~\cite{velickovic2017graph} to learn the dynamics of the influences, which shows superior performance against methods without this mechanism in Section~\ref{sec:exp:effect}. To the best of our knowledge, we are the first to use graph attentional network in the setting of traffic signal control.

\begin{figure}[t]
  \centering
  \includegraphics[width=.95\linewidth]{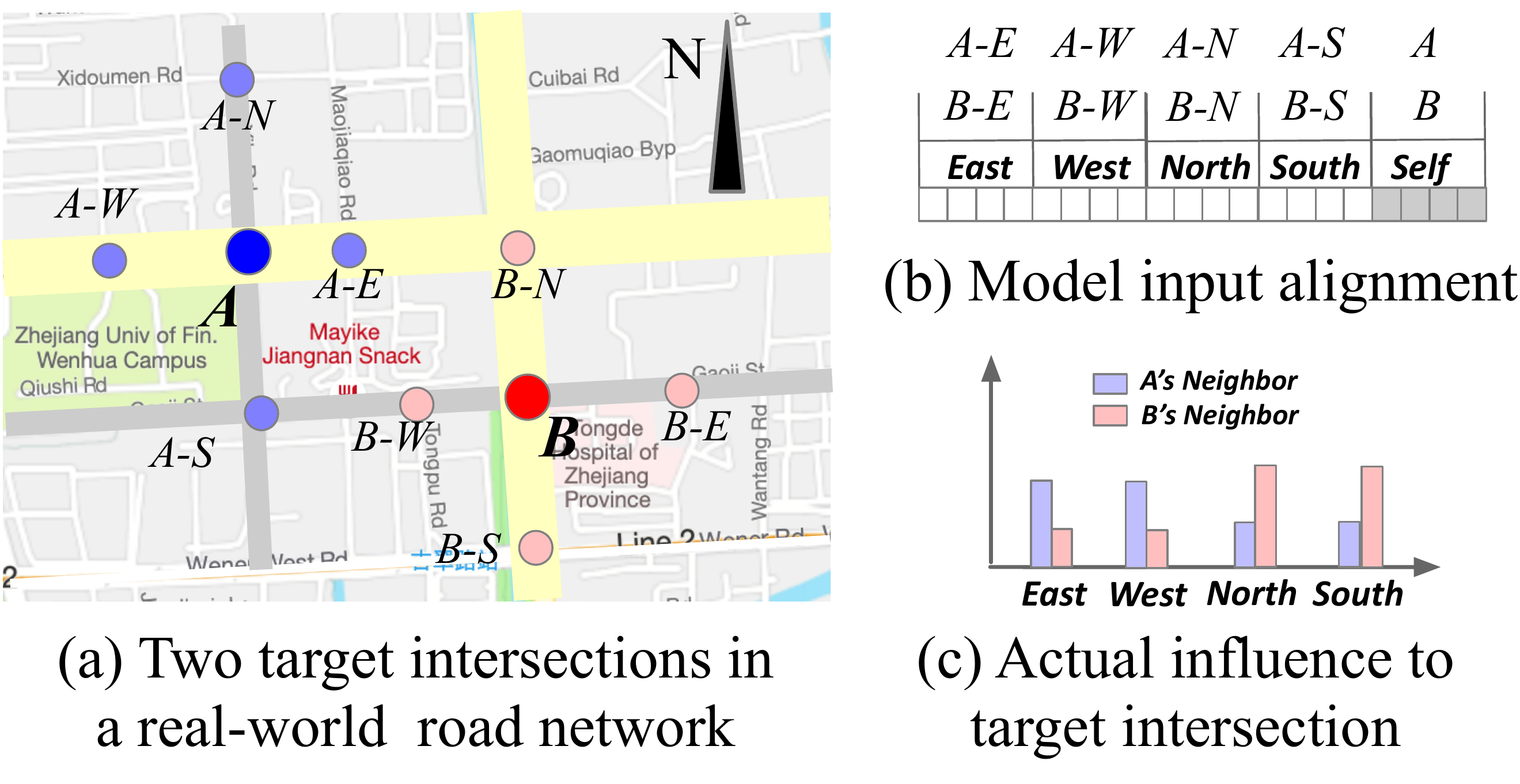}
  \caption{Illustration of index-based concatenation. 
  Thick yellow lines are the arterials and grey thin lines are the side streets. With index-based concatenation, \emph{A} and \emph{B}'s observation will be aligned as model inputs with an fixed order. These two inputs will confuse the model shared by \emph{A} and \emph{B}.}
  \label{fig:intro}
  \vspace{-3mm}
\end{figure}

$\bullet$ 
\emph{Index-free model learning with parameter sharing}. Agents often share the same model parameters to enable efficient learning~\cite{ALUK10,EAA13}. However, if we simply take neighboring information as part of the state for the communication purpose, the fixed indexing of neighbors may cause conflicting learning problems. Take Figure~\ref{fig:intro} as an example. Figure~\ref{fig:intro}(a) shows two target intersections \emph{A} and \emph{B}, where their neighbors \emph{A-W}, \emph{A-E}, \emph{B-N} and \emph{B-S} are on the major roads, while \emph{A-N}, \emph{A-S}, \emph{B-W} and \emph{B-E} are on the side streets. 
Intuitively, the two intersections on the major roads should relate more to the target intersection than those on the side streets, e.g., the influence from \emph{A-W} to \emph{A} should be greater than that from \emph{A-N} to \emph{A}. When neighbors are indexed in the way as [E, W, N, S], agent \emph{A} would care more about the first and second neighbors (i.e., East and West) and agent \emph{B} would pay more attention to the third and fourth neighbors (i.e., North and South). However, when two agents share the model parameters, this will cause conflicts to learn the influences of neighbors to the target intersection. 
In this paper, similar to the ``mean-field'' idea~\cite{yang2018mean}, we tackle this problem by averaging over the influences of all neighboring intersections with the learned attention weights, instead of using a fixed indexing for neighbors. This weighted average mechanism provides index-free modeling of the influences of neighboring intersections, and along with the parameter sharing strategy, the overall parameters of the learning model can be largely reduced. We will show the parameter analysis later in Section~\ref{sec:method:complex}.

\nop{
When concatenating the traffic condition into the target intersection's observation, existing methods tend to use parameter sharing (i.e, having all the intersections share parameters and maintain one model~\cite{ALUK10,EAA13}). This is because if we build models for every intersection, the number of parameters grows with the number of agents, which will take longer time for RL agents to converge. However, one important issue for methods with parameter sharing is that, they need to conduct an indexing mechanism to align the order of neighbors and yield a consistent input to the shared model. Otherwise, the inputs to the model will be inconsistent, making learning difficult~\cite{indexfree2010}. Take Figure~\ref{fig:intro} as an example. Figure~\ref{fig:intro}(a) shows two target intersections \emph{A} and \emph{B}, where their neighbors \emph{A-W}, \emph{A-E}, \emph{B-N} and \emph{B-S} are on the arterial, and \emph{A-N}, \emph{A-S}, \emph{B-W} and \emph{B-E} are on the side streets.
Intuitively, as indicated in Figure~\ref{fig:intro}(c), for intersection \emph{A}, neighbors on the arterial (\emph{A-W} and \emph{A-E}) have larger influence than the side-street neighbors (\emph{A-N} and \emph{A-S}). Similarly, for intersection \emph{B}, the influence of \emph{B-W} and \emph{B-E} is relatively smaller.

However, when \emph{A} and \emph{B}'s observation are aligned as model inputs with an order (``east'', ``west'', ``north'' and ``south'' in this case) which is fixed for all agents, these two inputs will confuse the model shared by \emph{A} and \emph{B} and the model is likely not able to differentiate the actual influence of neighbors to the target intersection.
In this paper, similar to the ``mean-field'' idea~\cite{yang2018mean}, we tackle this problem by averaging over the influences of all neighboring intersections with learned weights, instead of concatenating them all together. This weighted average mechanism provides index-free modeling of the influences of neighboring intersections, and along with the parameter sharing strategy, the overall parameters of the learning model can be largely reduced. We will show the parameter analysis later in Section~\ref{sec:method:complex}.
}

$\bullet$ 
\emph{Experiment on the large-scale road network}. To the best of our knowledge, none of the existing studies that use RL to cooperate traffic signals have evaluated their methods in large-scale road networks with hundreds of traffic signals. Instead, most of them justify their cooperation strategies on small road networks with only fewer than 20 intersections~\cite{wei2019survey}. In this paper, we conduct comprehensive experiments on both synthetic and real-world data, including a large-scale road network with 196 intersections derived from Manhattan, New York. The experiments demonstrate that the proposed model benefits from the dynamic communication and the index-free modeling mechanism and significantly outperforms the state-of-the-art methods.

\vspace{-1mm}
\section{Related Work}

Conventional coordinated methods~\cite{Manual} and systems~\cite{SCATS,SCOOT1,SCOOT2} in transportation usually coordinate traffic signals through modifying the offset (i.e., the time interval between the beginnings of green lights) between consecutive intersections and require the intersections to have the same cycle length. But this type of methods can only optimize the traffic flow for certain pre-defined directions~\cite{MAXBAND}. Actually, it is not an easy task to coordinate the offsets for traffic signals in the network. For network-level control, Max-pressure~\cite{Maxpressure} is a state-of-the-art signal control method which greedily takes actions that maximizes the throughput of the network, under the assumption that the downstream lanes have unlimited capacity. Other traffic control methods like TUC~\cite{TUC} also use optimization techniques to minimize vehicle travel time and/or the number of stops at multiple intersections under certain assumptions, such as the traffic flow is unform in a certain time period. However, such assumptions often do not hold in the network setting and therefore prevent these methods from being widely applied.

\nop{Conventional Transportation methods coordinate traffic signals among different intersections through modifying the offset (i.e. the time interval between phases at subsequent traffic signals). In a grid road network with homogeneous blocks, the coordination can be achieved through setting a fixed offset among all intersections. However, this could only provide coordination for certain direction and hinders the travel time of vehicles on other directions. Max-pressure method heuristically choose the phase that can maximize a pre-defined performance metric. It directly chooses to set green light for the phase with maximum ``pressure'', a pre-defined performance index.  However, such approaches still rely on assumptions to simplify the traffic condition, and do not guarantee optimal results in real world.
}

Recently, reinforcement learning techniques have been proposed to coordinate traffic signals for their capability of online optimization without prior knowledge about the given environment. One way to achieve coordination is through centralized optimization over the joint actions of multiple intersections.~\cite{PrBh11} directly trains one central agent to decide the actions for all intersections but it cannot learn well due to the curse of dimension in joint action space.~\cite{VaOl16,kuyer08} propose to jointly model two adjacent intersections and then using centralized coordination over the global joint actions, but they require the maximization over a combinatorially large joint action space and face scalability issues during deployment. Therefore, these centralized methods are hard to apply on the large-scale road network.

To mitigate this issue, independently modelling RL methods~\cite{Wier00,CaKr03,presslight} are proposed in which they train a bunch of RL agents separately, one for each intersection. 
To avoid the non-stationary impacts from neighboring agents,  communication among agents~\cite{Bish06,NoVr12} are proposed in order to achieve coordination using neighboring information: ~\cite{SBPB+06,EAA13,DrSt06,Wier00,chu2019multi} add downstream information into states,~\cite{ALUK10, Wier00} add all neighboring states, and~\cite{GCN18} adds neighbors' hidden states. However, in these methods, the information from different neighbors is concatenated all together and treated with equal importance, which leads to two major issues: 1). the impacts from neighboring intersections are changing dynamically with the traffic flow and should not be treated evenly. Even when the traffic flow is static, Kinenmatic-wave theory~\cite{hayes1970kinematic} from the transportation area shows that the upstream intersections could have larger influence than downstream intersections; 2). simple concatenation of neighboring information requires all agents to have an extra indexing mechanism, which is usually unrealistic and requires heuristic design. To address the shortcomings of previous methods, our proposed method leverages the attention mechanism to learn and specify different weights to different neighboring intersections and directly models the overall influence of neighboring intersections on the target intersection.

It is worth noting that most of joint modelling methods for the traffic signal control problem only conduct experiments on simple road networks with at most 20 intersections. To the best of our knowledge, none of the individual modelling methods conduct experiments on more than 70 signals~\cite{wei2019survey}. In this paper, we conduct experiments on the simulator under different scales, including a real-world road network with about 200 intersections. 


\nop{
What's more, current coordination only considers regional control up to 60 intersections because for a environment with large-scale agents, they usually need more neighbor information which as a result hinders their learning speed. our proposed methods coordinates the traffic signals with scalability, which is more suitable for city-wide control with hundreds of traffic signals. 
}

\vspace{-1mm}
\section{Problem Definition}

\begin{figure*}[t]
  \centering
  \includegraphics[width=.85\linewidth]{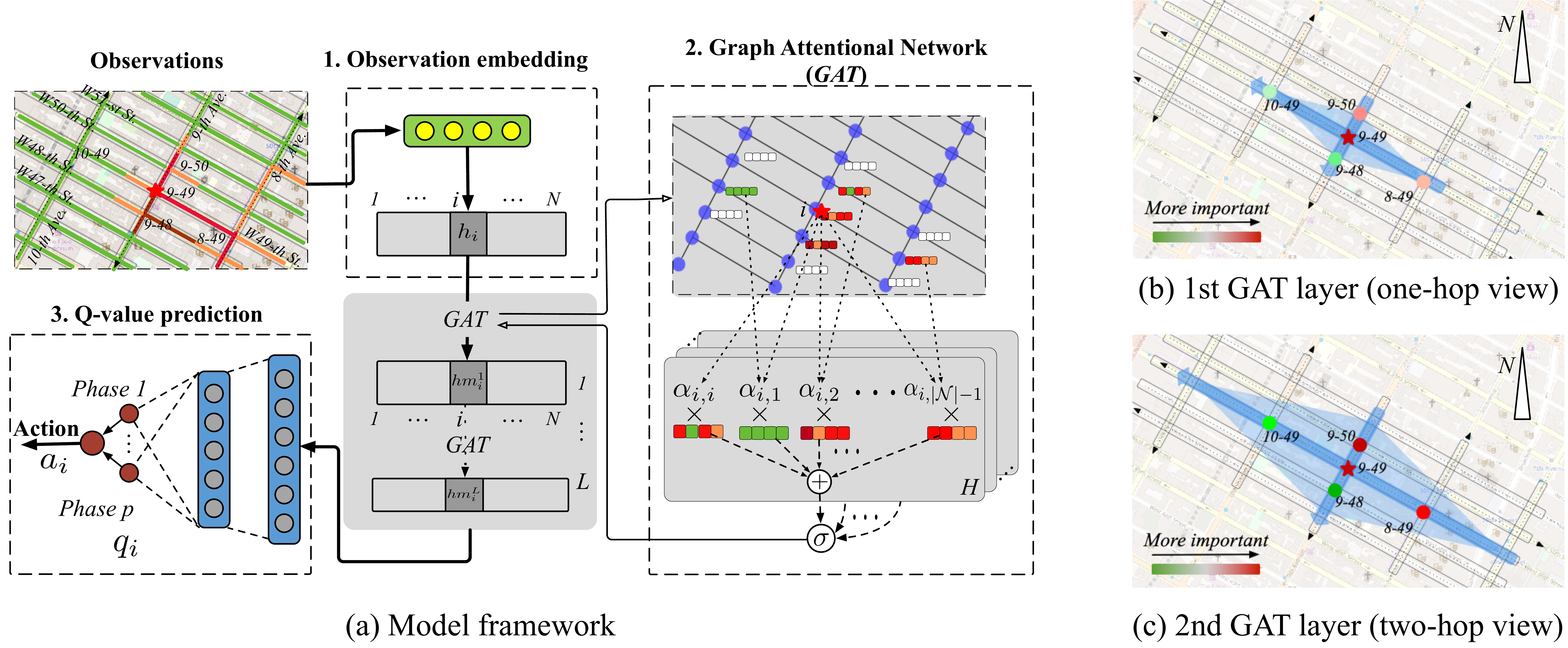}
  \caption{Left: Framework of the proposed \GAT model. Right: variation of cooperation scope (light blue shadow, from one-hop to two-hop) and attention distribution (colored points, the redder, the more important) of the target intersection.}
  \label{fig:framework}
  \vspace{-3mm}
\end{figure*}

In this section, we present the problem of traffic signal control as a Markov Game. Each intersection in the system is controlled by an agent. Given each agent observes part of the total system condition, we would like to proactively decide for all the intersections in the system which phases should they change to so as to minimize the average queue length on the lanes around the intersections.  Specifically, the problem is characterized by the following major components $\braket{\states,\observations,\actions,\transition, \reward,\policy,\decay}$:

$\bullet$ 
System state space $\states$ and observation space $\observations$. We assume that there are $N$ intersections in the system and each agent can observe part of the system state $\state\in\states$ as its observation $\observation \in \observations$. In this work, we define $\observation^t_i$ for agent $i$ at time $t$, which consists of its current phase (which direction is in green light) represented by a one-hot vector, and the number of vehicles on each lane connected with the intersection. 

$\bullet$ 
Set of actions $\actions$. In the traffic signal control problem, at time $t$, an agent $i$ would choose an action $a^t_i$ from its candidate action set  $\actions_i$ as a decision for the next $\Delta t$ period of time. Here, each intersection would choose a phase $p$ as its action $a^t_i$ from its pre-defined phase set, indicating that from time $t$ to $t+\Delta t$, this intersection would be in phase $p$.

$\bullet$
Transition probability $\transition$. Given the system state $\state^t$ and corresponding joint actions $\actionVec^t$ of agents at time $t$, the system arrives at the next state $\state^{t+1}$ according to the state transition probability $\transition(\state^{t+1}| \state^t,\actionVec^t): \states \times \actions_1 \times \dots \times \actions_N \rightarrow \Omega(\states)$, where $\Omega(\states)$ denotes the space of state distributions.

$\bullet$
Reward $\reward$. Each agent $i$ obtains an immediate reward $\reward^t_i$ from the environment at time $t$ by a reward function $\states \times \actions_1 \times \dots \times \actions_N \rightarrow \mathbb{R}$. In this paper, we want to minimize the travel time for all vehicles in the system, which is hard to optimize directly. Therefore, we define the reward for intersection i as $\reward^t_i = -\Sigma_l u^t_{i,l}$ where $u^t_{i,l}$ is the queue length on the approaching lane $l$ at time t. 

$\bullet$
Policy $\policy$ and discount factor $\decay$. Intuitively, the joint actions have long-term effects on the system, so that we want to minimize the expected queue length of each intersection in each episode. Specifically, at time $t$, each agent chooses an action following a certain policy $\observations\times \actions\rightarrow \policy$, aiming to maximize its total reward $G^t_i=\Sigma_{t=\tau}^T \decay^{t-\tau}\reward^t_i$, where $T$ is total time steps of an episode and $\decay\in[0,1]$ differentiates the rewards in terms of temporal proximity.

In this paper, we use the action-value function $Q_i(\theta_n)$ for each agent $i$ at the $n$-th iteration (parameterized by $\theta$)
to approximate the total reward $G^t_i$ with neural networks by minimizing the loss:
\begin{equation}
    \mathcal{L}(\theta_n)=\mathbb{E}[(\reward^t_i+\decay \max_{a'}Q({o^t_i}',{a^t_i}';\theta_{n-1})-Q(o^t_i,a^t_i;\theta_n))^2]\label{eq:bellman}
\end{equation}
where ${o^t_i}'$ denotes the next observation for $o^t_i$. These earlier snapshots of parameters are periodically updated with the most recent network weights and help increase the learning stability by decorrelating predicted and target q-values.

\section{Method}


In this section, we will first introduce the proposed cooperated RL network structure, as Figure~\ref{fig:framework} illustrates, from bottom to top layer: the first observation embedding layer, the interior neighborhood cooperation layers (shorted as GAT layers) and the final q-value prediction layer. Then we will discuss its time and space complexity compared with other methods of signal control for multiple intersections.
\subsection{Observation Embedding}
Given the raw data of the local observation, i.e., the number of vehicles on each lane and the phase the signal currently in,
we first embed such $k$-dimensional data into an $m$-dimensional latent space via a layer of Multi-Layer Perceptron:
\begin{equation}
h_i=Embed(o^t_i)=\sigma(o_iW_e+b_e),
\end{equation}
where $o^t_i\in\mathbb{R}^k$ is intersection $i$'s observation at time $t$
and $k$ is the feature dimension of $o^t_i$, $W_e\in\mathbb{R}^{k\times m}$ and $b_e\in\mathbb{R}^m$ are weight matrix and bias vector to learn, $\sigma$ is ReLU function (same denotation for the following $\sigma$). The generated hidden state $h_i\in\mathbb{R}^m$ represents the current traffic condition of the $i$-th intersection.

\subsection{Graph Attention Networks for Cooperation\label{sec:AoN}}
Communication between agents is necessary for cooperation in multi-agent reinforcement learning (MARL) environment, since the evaluation of the conducted policy for each agent depends not only on the observable surrounding, but also on the policies of other agents~\cite{NoVr12,de2009multiagent}. \GAT agent learns to communicate on the representations of neighboring intersections by leveraging the attention mechanism, a widely-used technique to boost model accuracy~\cite{yao2018modeling,you2016image,cheng2018neural,jiang2018graph}. Then an overall summary of the agent's neighborhood is generated, upon which the agent learns to model the influence of neighborhoods.

\subsubsection{Observation Interaction}
To learn the importance of information from intersection $j$ (source intersection) in determining the policy for intersection $i$ (target intersection), we first embed the representation of these two intersections from previous layer and calculate $e_{i,j}$, the importance of $j$ in determining the policy for $i$, with the following operation: 
\begin{equation}
e_{ij}=(h_iW_t)\cdot(h_jW_s)^T,
\end{equation}
where $W_s, W_t\in\mathbb{R}^{m\times n}$ are embedding parameters for the source and target intersection respectively.

Note that $e_{ij}$ is not necessarily equal to $e_{ji}$. Take the scenario in Figure~\ref{fig:framework}(a) as an example, where the 9-th Avenue is a one-way arterial. On one hand, since the traffic flow goes from \emph{Inter 9-50} to \emph{Inter 9-49}, the traffic condition from \emph{Inter 9-50} is important for \emph{Inter 9-49} to decide for the future actions, thus, $e_{9\text{-}49,9\text{-}50}$ should be quite large. On the other hand, as the traffic condition of the downstream \emph{Inter 9-49} is less helpful to \emph{Inter 9-50}, $e_{9\text{-}50,9\text{-}49}$ should be relatively small.
\subsubsection{Attention Distribution within Neighborhood Scope}
\label{sec:neighbor-scope}

To retrieve a general attention value between source and target intersections, we further normalize the interaction scores between the target intersection $i$ and its neighborhood intersections:
\begin{equation}
\alpha_{ij}=\text{softmax}(e_{ij})=\frac{\exp(e_{ij}/\tau)}{\sum_{j\in\mathcal{N}_i}\exp(e_{ij}/\tau)},
\end{equation}
where $\tau$ is the temperature factor
and $\mathcal{N}_i$ is the set of intersections in the target intersection's \textit{neighborhood scope}. The neighborhood of the target contains the top $|\mathcal{N}_i|$ closest intersections, and the distance can be defined in multiple ways. For example, we can construct the neighborhood scope for target intersection $i$ through:

$\bullet$
\emph{Road distance}: the geo-distance between two intersections' geo-locations.

$\bullet$
\emph{Node distance}: the smallest hop count between two nodes over the network, with each node as an intersection.

Note that intersection $i$ itself is also included in $\mathcal{N}_i$ to help the agent get aware of how much attention should be put on its own traffic condition.



The general attention score $\alpha_{ij}$ is beneficial not only for it applies to all kinds of road network structures (intersections with different numbers of arms), but also for it relaxes the concept of ``neighborhood''. Without losing generality, the target can even take some other intersections into $\mathcal{N}_i$ although they are not adjacent to them. For instance, one four-way intersection can determine its signal policy based on information from five nearby intersections, four of which are the adjacent neighbors while the other is disconnected but geographically close to the target intersection. 

\subsubsection{Index-free Neighborhood Cooperation}
To model the overall influence of neighborhoods to the target intersection, the representation of several source intersections are combined with their respective importance:
\begin{equation}
hs_i=\sigma\big(W_q\cdot\sum_{j\in\mathcal{N}_i}\alpha_{ij}(h_jW_c)+b_q\big),
\end{equation}
where $W_c\in\mathbb{R}^{m\times c}$ is weight parameters for source intersection embedding, $W_q$ and $b_q$ are trainable variables. The weighted sum of neighborhood representation $hs_i\in\mathbb{R}^c$ accumulates the key information from the surrounding environment for performing efficient signal policy. By summing over the neighborhood representation, the model is \emph{index-free} that does not require all agents to align the index of their neighboring intersections. 

The graph-level attention allows the agent to adjust their focus according to the dynamic traffic and to sense the environment in a larger scale. In Figure~\ref{fig:framework}(b) and (c), the emphasizes of \emph{Inter 9-49} on four neighbors are quite distinct due to the uni-directional traffic flow, i.e., a higher attention score for \emph{Inter 9-50} (upstream, red marked) than for \emph{Inter 9-48} (downstream, green marked).
The agent for \emph{Inter 9-49} acquires the knowledge of adjacent intersections (\emph{Inter 9-48}, \emph{Inter 9-50}, \emph{Inter 10-49} and \emph{Inter 8-49}) directly from the first layer of Graph Attention Networks (GAT). Since the hidden states of adjacent neighbors from the first GAT layer carry their respective neighborhood message, then in the second GAT layer, the cooperation scope of \emph{Inter 9-49} expands significantly (blue shadow in Figure~\ref{fig:framework}(c)) to 8 intersections. Such additional information helps the target \emph{Inter 9-49} learn the traffic trend. As a result, \emph{Inter 9-49} relies more on the upstream intersections and less on the downstream to take actions, and the attention scores on \emph{Inter 9-50} and \emph{Inter 8-49} grow higher while those on \emph{Inter 10-49} and \emph{Inter 9-48} become lower. More GAT layers helps the agent detect environment dynamics more hops away.

\subsubsection{Multi-head Attention}
The cooperating information $hs_i$ for the $i$-th intersection concludes one type of relationship with neighboring intersections. To jointly attend to the neighborhood from different representation subspaces at different positions, we extend the previous single-head attention in the neural network to multi-head attention as much recent work did~\cite{vaswani2017attention,velickovic2017graph}. Specifically, the attention function (procedures including \emph{Observation Interaction}, \emph{Attention Distribution} and \emph{Neighborhood Cooperation}) with different linear projections (multiple sets of trainable parameters \{$W_t$, $W_s$, $W_c$\}) is performed in parallel and the different versions of neighborhood condition summarization $hs_i$ are averaged as $hm_i$:
\begin{equation}
    e_{ij}^h=(h_iW_t^h)\cdot(h_jW_s^h)^T
\end{equation}
\vspace{-4mm}
\begin{equation}
   \alpha_{ij}^h=\text{softmax}(e_{ij}^h)=\frac{\exp(e_{ij}^h/\tau)}{\sum_{j\in\mathcal{N}_i}\exp(e_{ij}^h/\tau)}
   \label{eq:multi-head_score}
\end{equation}
\vspace{-4mm}
\begin{equation}
   hm_i=\sigma\Big(W_q\cdot\big(\frac{1}{H}\sum_{h=1}^{h=H}\sum_{j\in\mathcal{N}_i}\alpha_{ij}^h(h_jW_c^h)\big)+b_q\Big)
   \label{eq:multi-head_attention}
\end{equation}
where $H$ is the number of attention heads. 
Besides averaging operation, concatenating the product of multi-head attention is another feasible way to conclude multiple types of the neighborhood cooperation.

In this work, we investigate the effects of multi-head attention on performance of the proposed model and find that $5$ attention heads achieve the best performance.




\subsection{Q-value Prediction}
As illustrated in Figure~\ref{fig:framework}(a), each hidden layer of model \GAT learns the neighborhood representation through methods introduced in Section~\ref{sec:AoN}. We denote such layerwise cooperation procedure by \emph{GAT}, then the forward propagation of input data in \GAT can be formatted as follows:

\begin{equation}
    \begin{aligned}
    hi&=&Embed(o^t_i),\\
    hm_i^1&=&GAT^1(h_i),\\
    &\cdots&,\\
    hm_i^L&=&GAT^L(hm_i^{L-1}),\\
    \widetilde{q}(o^t_i)&=&hm_i^LW_p+b_p,
    \end{aligned}
\end{equation}

where $W_p\in\mathbb{R}^{c\times p}$ and $b_p\in\mathbb{R}^p$ are parameters to learn, $p$ is the number of phases (action space), $L$ is the number of GAT layers, $\widetilde{q}$ is the predicted q-value.

According to Eq.~\eqref{eq:bellman}, the loss function for our \GAT to optimize the current policy is:
\begin{equation}
L(\theta)=\frac{1}{T}\sum_{t=1}^{t=T}\sum_{i=1}^{i=N}\big(q(o^t_i,a^t_i)-\widetilde{q}(o^t_i;a^t_i,\theta)\big)^2,
\end{equation}
where $T$ is the total number of time steps that contribute to the network update, $N$ is the number of intersections in the whole road network, $\theta$ represents all the trainable variables in \GAT.

As the importance of neighborhood to the target intersection varies spatially and temporally, the proposed attention mechanism is able to help the target agent distinguish among the complex scenarios by considering the traffic condition of any source-target intersection pair. 

\subsection{Complexity Analysis\label{sec:complexity_analysis}}
\label{sec:method:complex}
Although \GAT spares additional parameters to learn the dynamic cooperation from neighborhood, owing to the index-free parameter sharing mechanism, both the time and space it demands are approximately equal to $O(m^2L)$, which is irrelevant to the number of intersections. Hence \GAT is scalable even if the road network contains hundreds of or even thousands of intersections.

\subsubsection{Space Complexity\label{sec:space_complexity}}
If there are $L$ hidden layers and each layer has $m$ neurons, then the size of the weight matrices and bias vectors in each component of \GAT is: 1) \emph{Observation Embedding} layer: $km+m$; 2) interior \emph{Graph Attentional} layers: $\big(3m^2+(m^2+m)\big)L=m(4m+1)L$; 3) \emph{Q-value Prediction} layer: $mp+p$. Hence the total number of learnable parameters to store is $O\big(m(4mL+L+k+1+p)+p\big)$. Normally, the size of the hidden layer ($m$) is far greater than the number of layers ($L$), the phase space ($p$) and comparable to the input dimension ($k$). Therefore, the space complexity of \GAT is approximately equal to $O(m^2L)$.

If we leverage $N$ separate RL models (without parameter sharing) to control signals in $N$ intersections, then the space complexity is $O\Big(\big((km+m)+(m^2+m)L+(mp+p)\big)\cdot N\Big)\approx O(m^2L\cdot N)$, which is unfeasible when $N$ is extremely large for city-level traffic signal control. To scale up, the simplest solution is to allow all the intersections to share parameters and maintain one model, in this case, the space complexity is $O(m^2L)$, which is identical to that of \GAT.

\begin{figure*}[t!]
  \centering
  \begin{tabular}{ccc}
  \includegraphics[width=0.20\textwidth]{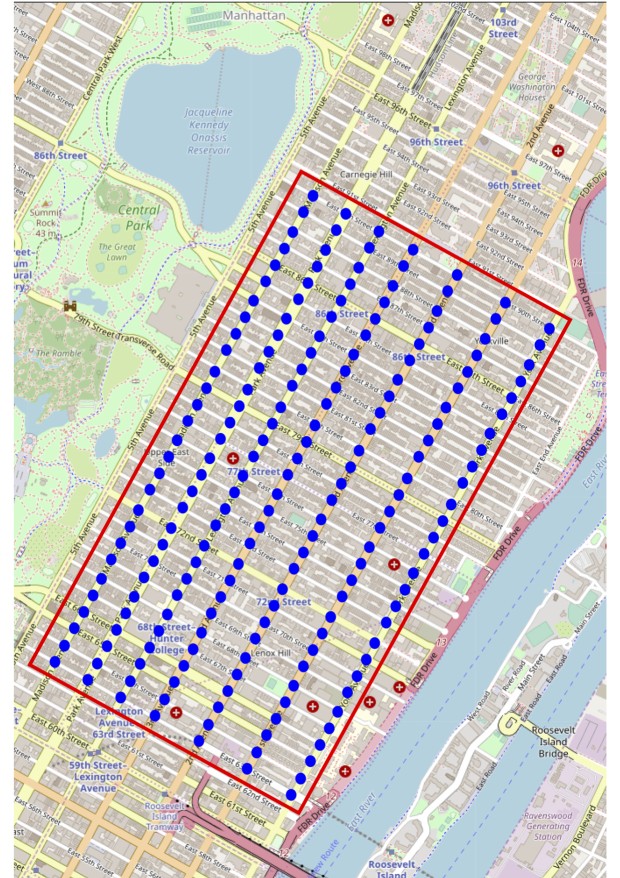}&
  \includegraphics[width=0.38\textwidth]{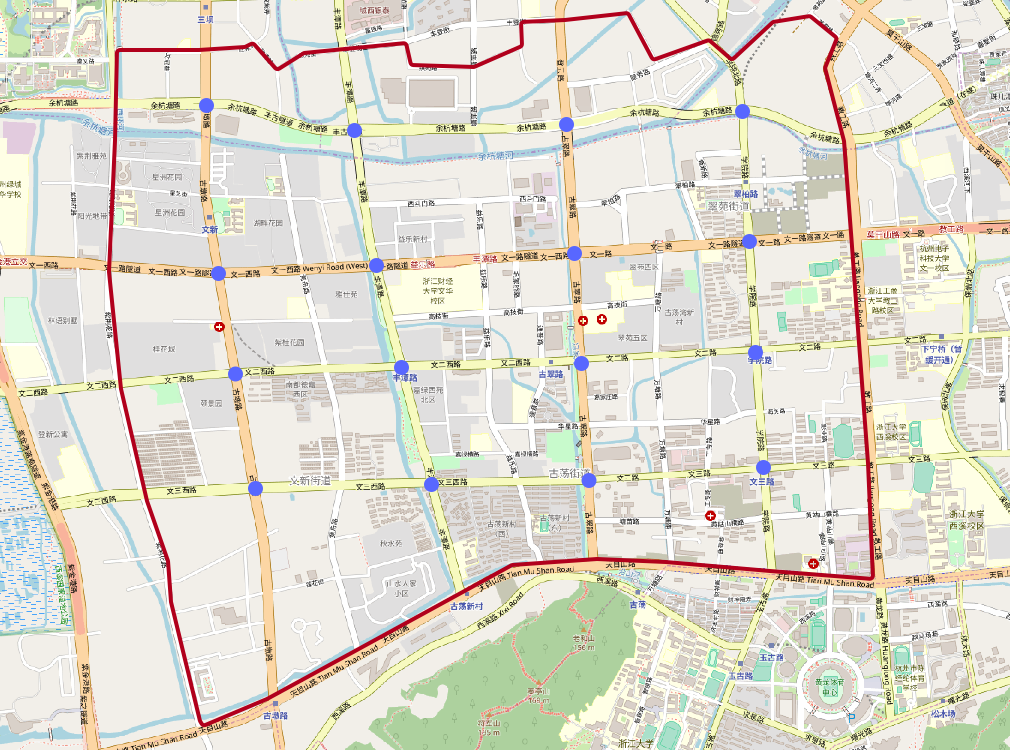}&
  \includegraphics[width=0.178\textwidth]{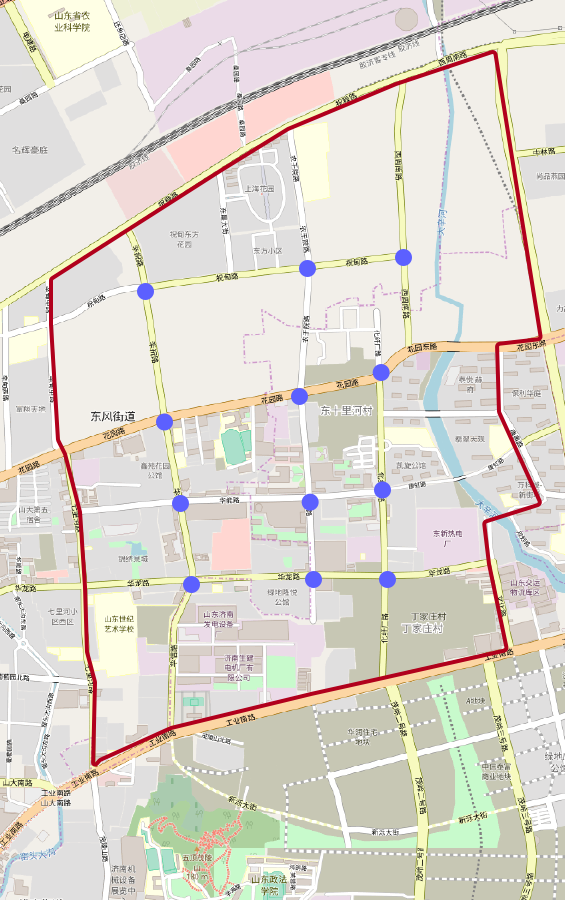}\\
  \begin{tabular}[c]{@{}c@{}}(a) Upper East Side, Manhattan,\\New York, USA\end{tabular} &
  \begin{tabular}[c]{@{}c@{}}(b) Gudang Sub-district,\\ Hangzhou, China\end{tabular}
  
  &
  \begin{tabular}[c]{@{}c@{}}(c) Dongfeng Sub-district,\\ Jinan, China\end{tabular}\\
  \end{tabular}
  \caption{Road networks for real-world datasets. Red polygons are the areas we select to model, blue dots are the traffic signals we control. Left: 196 intersections with uni-directional traffic, middle: 16 intersections with uni- \& bi-directional traffic, right: 12 intersections with bi-directional traffic.}
  \label{fig:real_road_net}
    \vspace{-3mm}
\end{figure*}

\subsubsection{Time Complexity\label{sec:time_complexity}}
We assume that: 1) all the agents leverage \GAT to predict q-values for the corresponding intersections concurrently; 2) the multiple heads of attention are independently computed so that they are as fast as the single-head attention; 3) the embeddings for either source or target intersection condition via $W_s$, $W_c$ and $W_t$ are separate processes that can also be executed at the same time, 4) for one target intersection, the interaction with all the neighbors is computed simultaneously, then the time complexity (only multiplication operations considered since the addition procedures are relatively insignificant) in each component of \GAT is: 1) \emph{Observation Embedding} layer: $km$; 2) interior \emph{Graph Attentional} layers: $(m^2+m^2)L$; 3) \emph{Q-value Prediction} layer: $mp$. Hence the time complexity is $O\big(m(k+2mL+p)\big)$, and similarly, it is approximately equal to $O(m^2L)$. 

Either the individual RL models or the shared single RL model for signal control in multiple intersections requires $O\big(m(k+mL+p)\big)\approx O(m^2L)$ computation, approaching that of \GAT.

\section{Experiments}
We perform experiments on two synthetic datasets and three real-world datasets to evaluate our  proposed method, especially the graph-level attention for neighborhood cooperation.

\nop{
We aim to answer the following research questions:

$\bullet$
 \textbf{RQ1}: Compared with state-of-the-art transportation and RL methods, how does \GAT perform?

$\bullet$
\textbf{RQ2}: Can \GAT scale up to large road networks in experiments with its time complexity in consistent with the theoretical complexity analysis in Section~\ref{sec:method:complex}?

$\bullet$
\textbf{RQ3}: How do different components (i.e., neighborhood definition, number of neighbors, and number of attention heads) affect \GAT?


$\bullet$
\textbf{RQ4}: Is the proposed model able to learn the spatial and temporal difference of influence?
}

\subsection{Settings}
We conduct experiments on CityFlow\footnote{http://cityflow-project.github.io}, an open-source traffic simulator that supports large-scale traffic signal control~\cite{huichu19}. After the traffic data being fed into the simulator, a vehicle moves towards its destination according to the setting of the environment. The simulator provides the state to the signal control method and executes the traffic signal actions from the control method. Following the tradition, each green signal is followed by a three-second yellow signal and two-second all red time.\footnote{\GAT's codes, parameter settings, public datasets can be found at: \url{https://github.com/wingsweihua/colight}. More datasets can be found at: \url{http://traffic-signal-control.github.io}}

In a traffic dataset, each vehicle is described as $(o, t, d)$, where $o$ is the origin location, $t$ is  time, and $d$ is the destination location. Locations $o$ and $d$ are both locations on the road network. 

\subsection{Datasets}
\subsubsection{Synthetic Data}
In the experiment, we use two kinds of synthetic data, i.e., uni- and bi-directional traffic, on the following different road networks:
~\noindent\\$\bullet$
\arterial: A $1\times 3$ arterial to show the spatial attention distribution learned by \GAT.
~\noindent\\$\bullet$
\gridthree: A $3\times 3$ grid network to show convergence speed of different RL methods and the temporal attention distribution.
~\noindent\\$\bullet$
\gridsix: A $6\times 6$ grid network to evaluate effectiveness and efficiency of different methods.

Each intersection in the synthetic road network has four directions (West$\rightarrow$East, East$\rightarrow$West, South$\rightarrow$North, North$\rightarrow$South), and 3 lanes (300 meters in length and 3 meters in width) for each direction. In bi-directional traffic, vehicles come uniformly with 300 vehicles/lane/hour in West$\leftrightarrow$East direction and 90 vehicles/lane/hour in South$\leftrightarrow$North direction. Only West$\rightarrow$East and  North$\rightarrow$South directional flows travel in uni-directional traffic. 
\subsubsection{Real-world Data}
We also use the real-world traffic data from three cities: New York, Hangzhou and Jinan. Their road networks are imported from OpenStreetMap\footnote{https://www.openstreetmap.org}, as shown in Figure~\ref{fig:real_road_net}. And their traffic flows are processed from multiple sources, with data statistics listed in Table~\ref{tab:real-dataset}. The detailed descriptions on how we preprocess these datasets are as follows:

\begin{table}[h!]
\caption{Data statistics of real-world traffic dataset}
\label{tab:real-dataset}
\begin{tabular}{lccccc}
\toprule
\multirow{2}{*}{Dataset} & \multirow{2}{*}{\# intersections} & \multicolumn{4}{c}{Arrival rate (vehicles/300s)} \\
                         &                                   & Mean         & Std         & Max       & Min       \\\midrule
\newYork                 & 196                                & 240.79       & 10.08       & 274       & 216       \\
\hangzhou                    & 16                                & 526.63       & 86.70       & 676   &256   \\
\jinan                 & 12                                & 250.70       & 38.21       & 335       & 208       \\\bottomrule
\end{tabular}
\vspace{-4mm}
\end{table}

~\noindent$\\\bullet$
\newYork: There are 196 intersections in Upper East Side of Manhattan with open source taxi trip data\footnote{http://www.nyc.gov/html/tlc/html/about/trip\_record\_data.shtml}. Since the taxi data only contains the origin and destination geo-locations of each trip, we first map these geo-locations to the intersections and find the shortest path between them. Then we take the trips that fall within the selected areas.
~\noindent$\\\bullet$
\hangzhou: There are 16 intersections in Gudang Sub-district with traffic data generated from the roadside surveillance cameras. Each record in the dataset consists of time, camera ID and the information about vehicles. By analyzing these records with camera locations, the trajectories of vehicles are recorded when they pass through road intersections. We use the number of vehicles passing through these intersections as traffic volume for experiments. 
~\noindent$\\\bullet$
\jinan: Similar to \hangzhou, this traffic data is collected by roadside cameras near 12 intersections in Dongfeng Sub-district, Jinan, China.





\begin{table*}[t!]
\centering
\caption{Performance on synthetic data and real-world data w.r.t average travel time. \GAT is the best.}\label{tab:overall_performance}
\begin{tabular}{lccccc}
\toprule
Model&\gridsix-Uni&\gridsix-Bi&\newYork&\hangzhou&\jinan\\\midrule
\formula~\cite{Manual}&209.68&209.68&1950.27&728.79&869.85\\
\maxPressure~\cite{Maxpressure}&186.07&194.96&1633.41&422.15&361.33\\
\midrule
\nips~\cite{VaOl16}&1532.75&2884.23&2187.12&1582.26&1210.70\\
\individualRL~\cite{wei2018intellilight}&314.82&261.60&-$^{*}$&345.00&325.56\\
\oneModel~\cite{chu2019multi}&181.81&242.63&1973.11&394.56&728.63\\
\oneModelNeighbor~\cite{ALUK10}&240.68&248.11&2280.92&1053.45&1168.32\\
\GCN~\cite{GCN18}&205.40&272.14&1876.37&768.43&625.66\\\midrule
\GATNode&178.42& 176.71&1493.37&331.50&340.70\\
\GATEuclidean&\bf 173.79&\bf170.11&\bf 1459.28&\bf 297.26&\bf 291.14\\\bottomrule
\end{tabular}
\\\footnotesize{$^{*}$No result as \individualRL can not scale up to 196 intersections in New York's road network.}
\vspace{-2mm}
\end{table*}

\subsection{Compared Methods}
We compare our model with the following two categories of methods: conventional transportation methods and RL methods. Note that all the RL models are learned without any pre-trained parameters for fair comparison.

\textbf{Transportation Methods:} 
~\noindent\\$\bullet$
\formula~\cite{Manual}: Fixed-time with random offsets. This method uses a pre-determined plan for cycle length and phase time, which is widely used when the traffic flow is steady.
~\noindent\\$\bullet$
\maxPressure~\cite{Maxpressure}: A state-of-the-art network-level traffic signal control method in the transportation field, which greedily chooses the phase that maximizes the pressure (a pre-defined metric about upstream and downstream queue length).

\textbf{RL Methods:}
~\noindent\\$\bullet$
\nips~\cite{VaOl16}: A RL-based method for multi-intersection signal control with joint-action modelling~\cite{VaOl16}. Specifically, the cooperation is achieved by designing a coordination graph and it learns to optimize the joint action between two intersections.
~\noindent\\$\bullet$
\individualRL~\cite{wei2018intellilight}: An individual deep RL approach which does not consider neighbor information. Each intersection is controlled by one agent, and the agents do not share parameters, but update their own networks independently.
~\noindent\\$\bullet$
\oneModel~\cite{chu2019multi}: This method uses the same state and reward as \individualRL in its agent design, which only considers the traffic condition on the roads connecting with the controlled intersection. Instead of maintaining their own parameters, all the agents share the same policy network.
~\noindent\\$\bullet$
\oneModelNeighbor~\cite{ALUK10}: Based on \oneModel, agents concatenate their neighboring intersections' traffic condition with their own and all the agents share the same parameters. Hence its feature space for observation is larger than \oneModel.
~\noindent\\$\bullet$
\GCN~\cite{GCN18}: A RL based traffic signal control method that uses a graph convolutional neural network to automatically extract the traffic features of adjacent intersections. This method treats each neighboring traffic condition without difference.

 \textbf{Variants of Our Proposed Method:}
~\noindent\\$\bullet$
\GATEuclidean: The neighborhood scope of an intersection is constructed through geo-distance.
~\noindent\\$\bullet$
\GATNode: The neighborhood scope is constructed through node distance, i.e., the smallest hop count between two intersections in the road network.

\subsection{Evaluation Metric}
Following existing studies~\cite{wei2018intellilight,presslight}, we use the average \textbf{travel time} to evaluate the performance of different models for traffic signal control. It calculates the average travel time of all the vehicles spend between entering and leaving the area (in seconds), which is the most frequently used measure of performance to control traffic signal in the transportation field~\cite{roess2004traffic,wei2019survey}.

\begin{figure*}[t!]
  \centering
  \subfigure[\gridthree]{\label{fig:convergence_3x3} \includegraphics[width=0.27\textwidth]{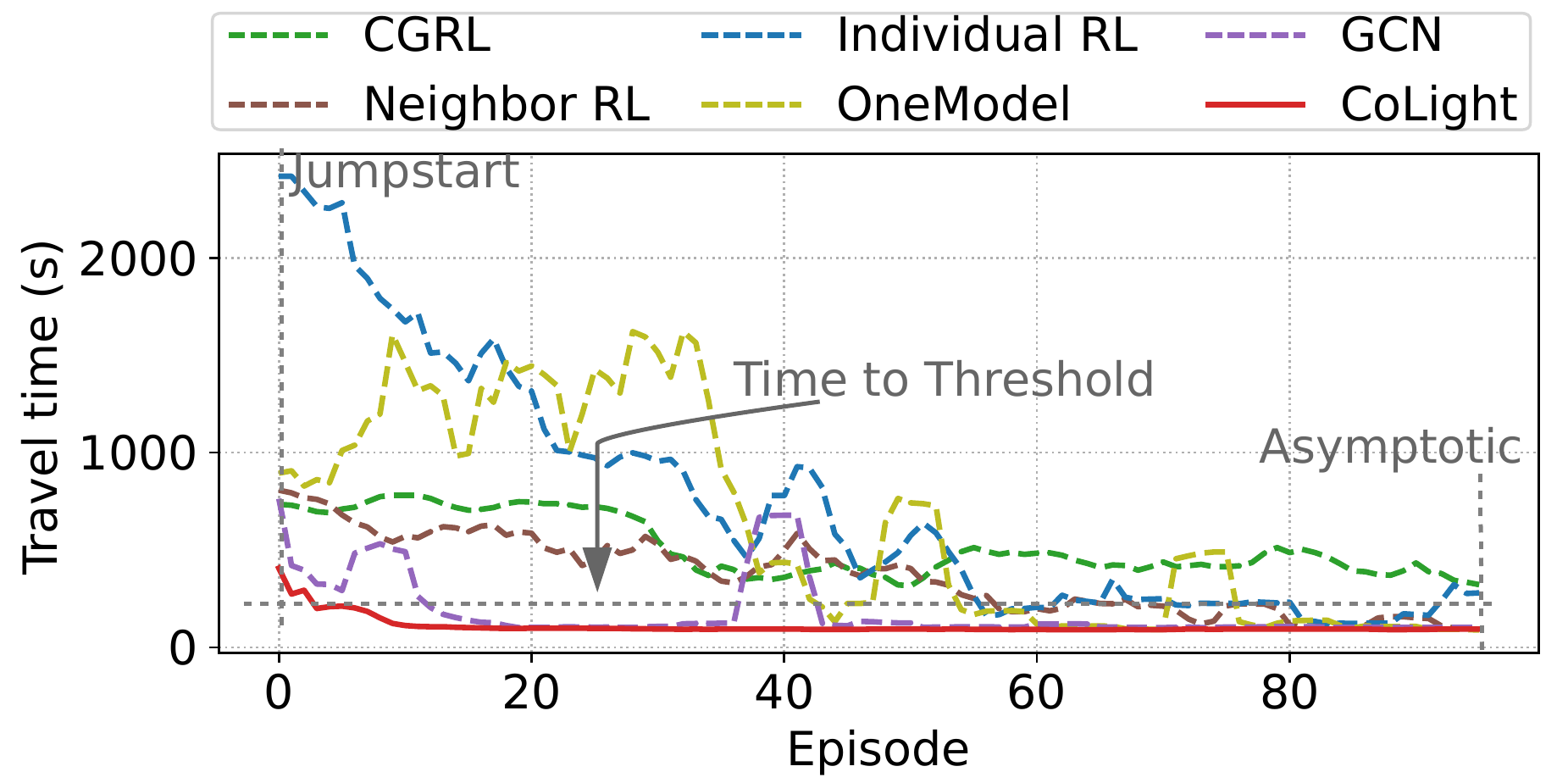}}
  \subfigure[\gridsix-Uni]{\label{fig:convergence_6x6_uni} \includegraphics[width=0.27\textwidth]{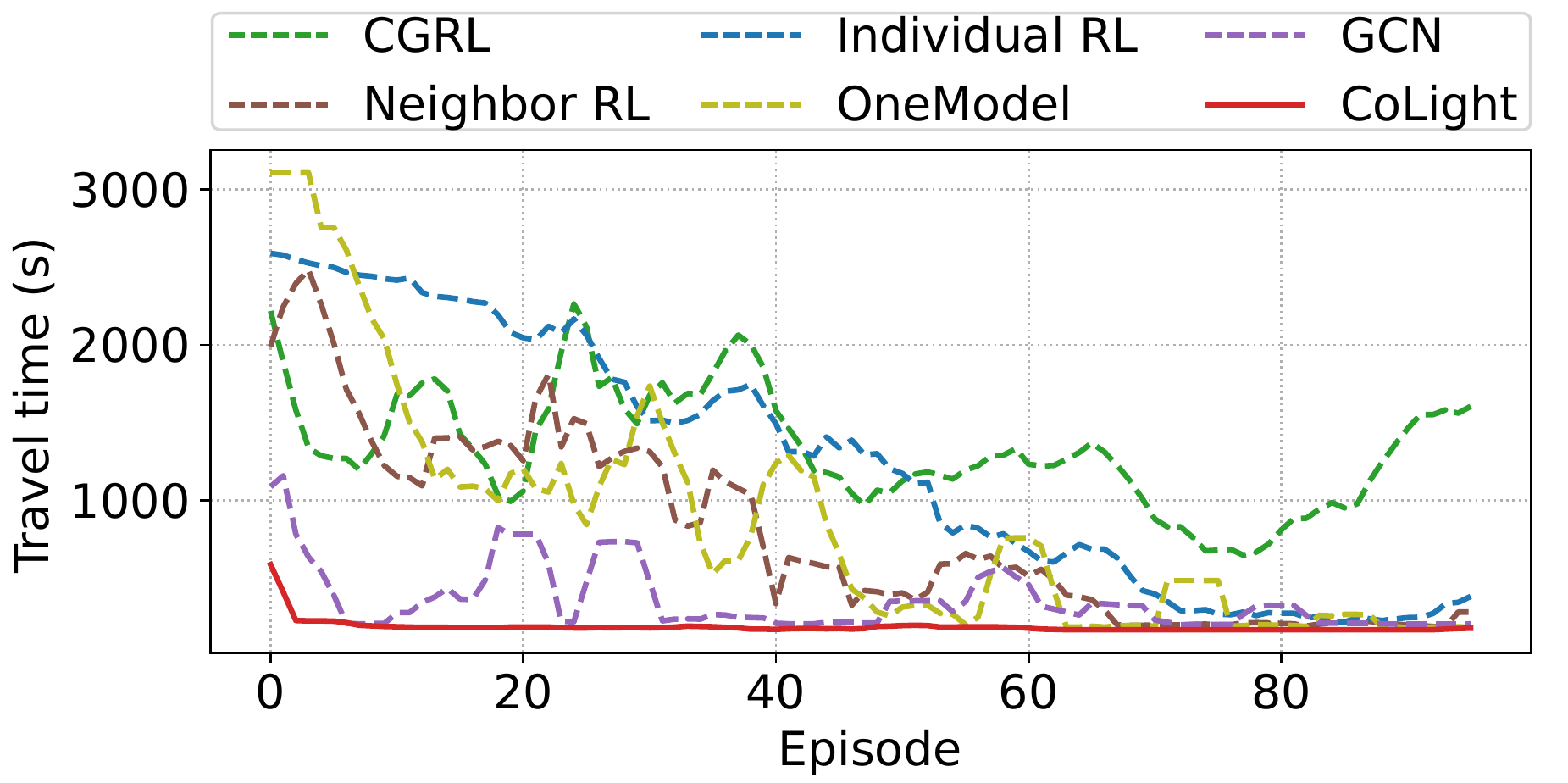}} 
  \subfigure[\gridsix-Bi]{\label{fig:convergence_6x6_bi} \includegraphics[width=0.27\textwidth]{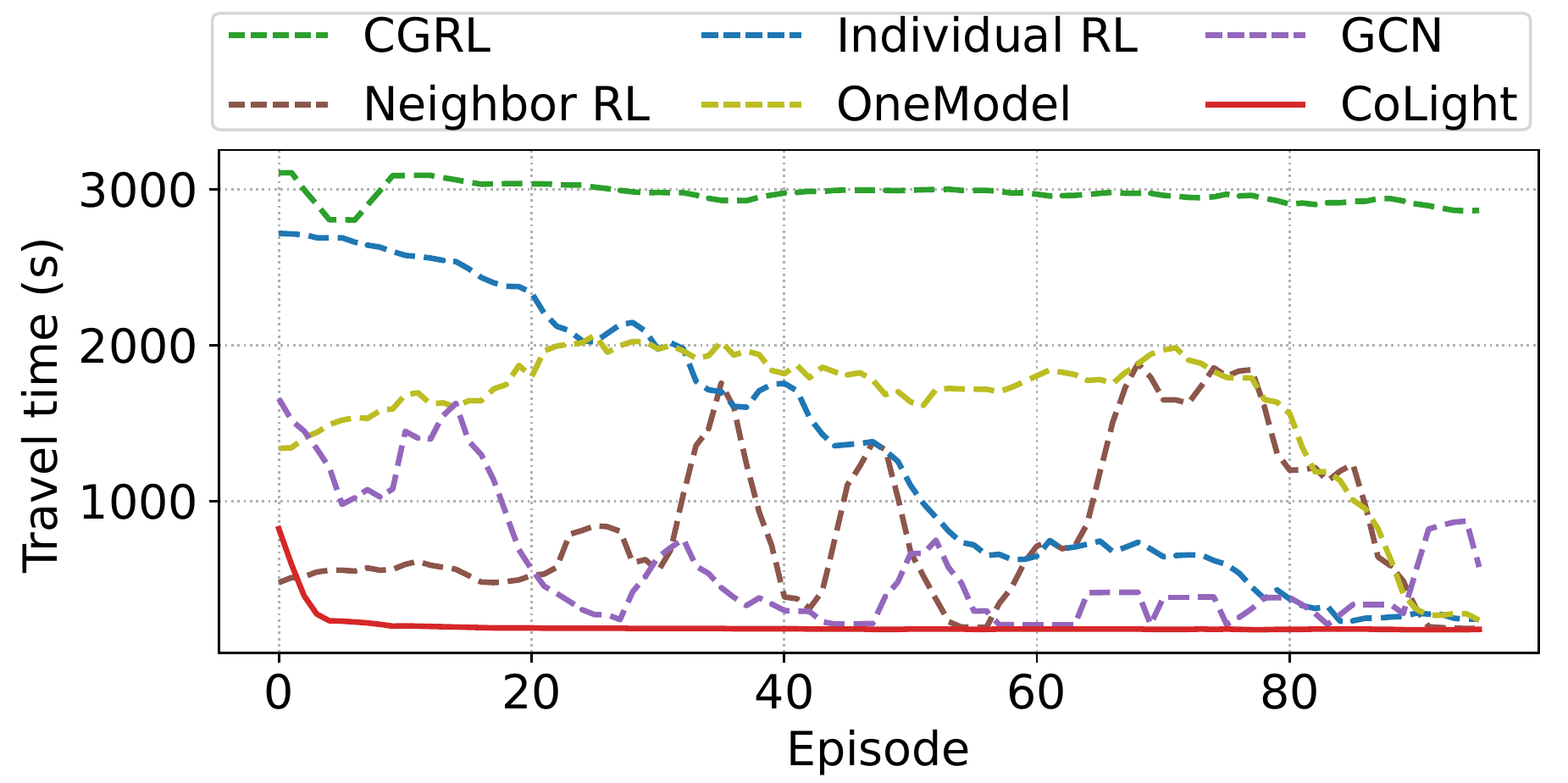}}  
  \subfigure[\newYork]{\label{fig:convergence_NewYork} \includegraphics[width=0.27\textwidth]{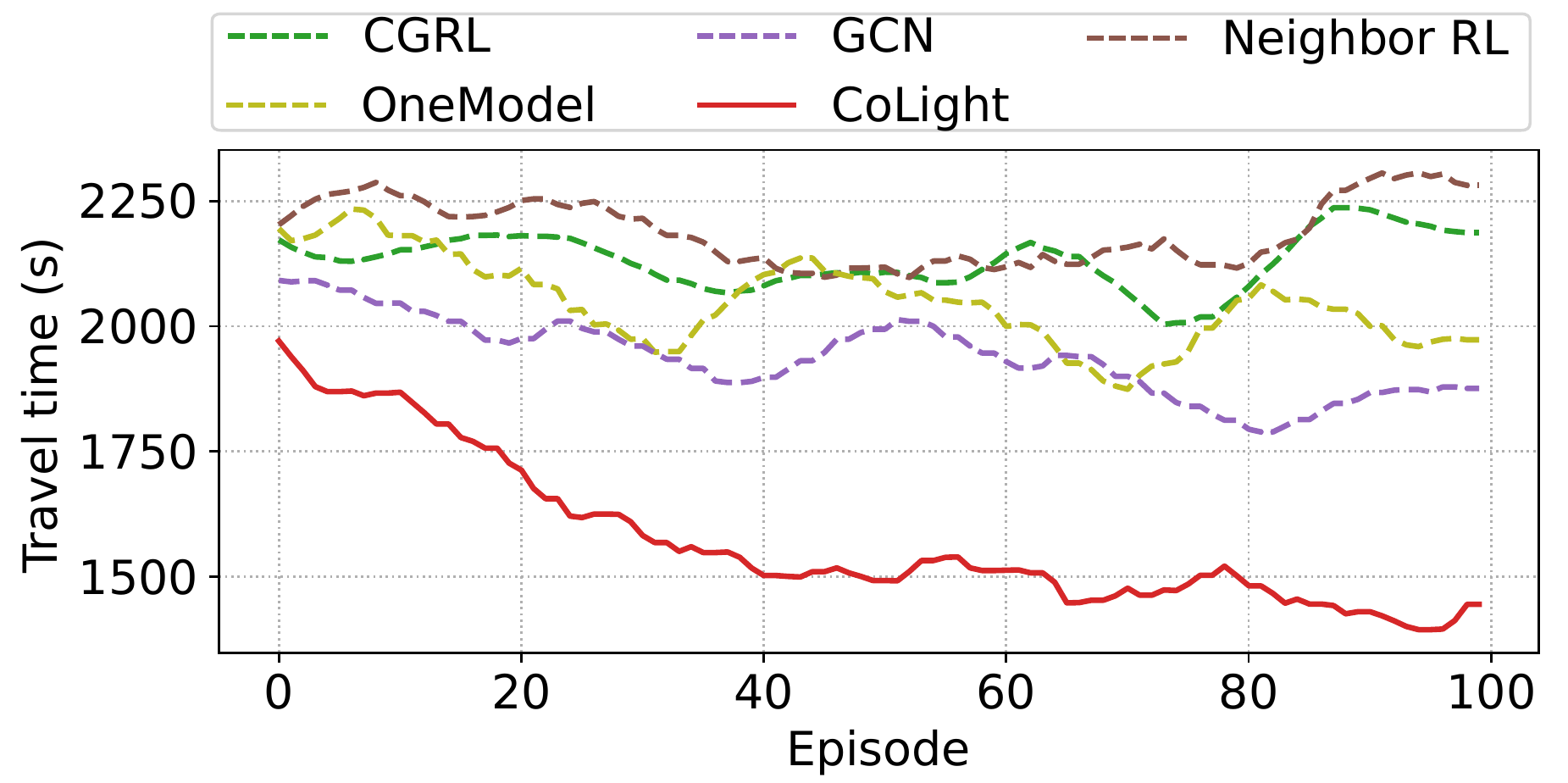}}
  \subfigure[\hangzhou]{\label{fig:convergence_Hangzhou} \includegraphics[width=0.27\textwidth]{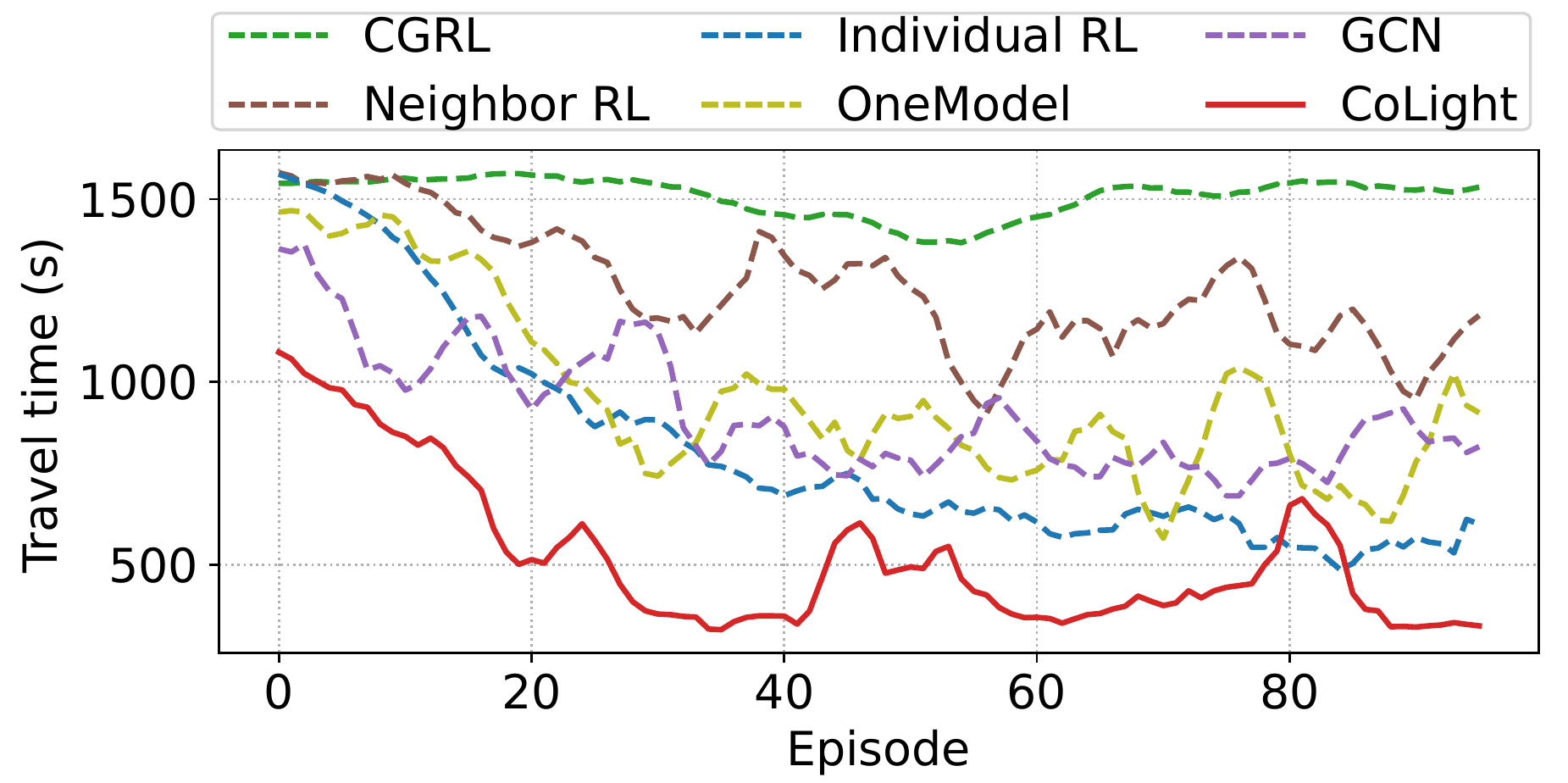}}
  \subfigure[\jinan]{\label{fig:convergence_Jinan} \includegraphics[width=0.27\textwidth]{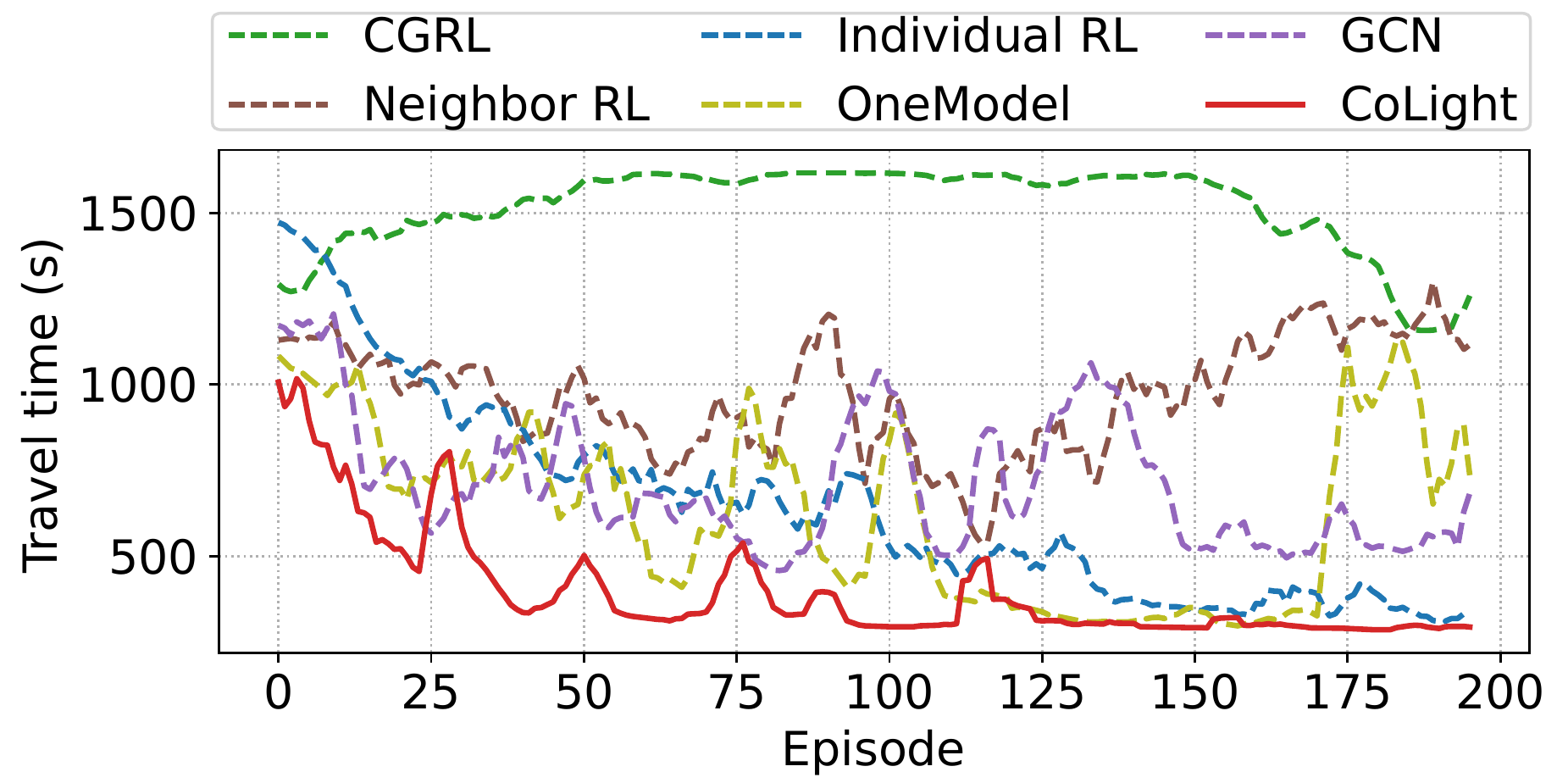}}
  \caption{Convergence speed of \GAT 
  (red continuous curves) 
  and other 5 RL baselines 
  (dashed curves) 
  during training. \GAT starts with the best performance (Jumpstart), reaches to the pre-defined performance the fastest (Time to Threshold), and ends with the optimal policy (Aysmptotic). Curves are smoothed with a moving average of 5 points.}
  \label{fig:convergence}
 \vspace{-3mm}
\end{figure*}

\subsection{Performance Comparison}
In this section, we investigate on the performance of \GAT w.r.t. the travel time and compare it with state-of-the-art transportation and RL methods. 
\label{sec:exp:effect}
\subsubsection{Overall Analysis}
Table~\ref{tab:overall_performance} lists the performance of two types of the proposed \GAT, classic transportation models as well as state-of-the-art learning methods in both synthetic and real-world datasets. We have the following observations:

1. \GAT achieves consistent performance improvements over state-of-the-art transportation (\maxPressure) and RL (\individualRL) methods across diverse road networks and traffic patterns:
the average improvement is $6.98\%$ for synthetic data and $11.69\%$
for real-world data.
The performance improvements are attributed to the benefits from dynamic communication along with the index-free modeling. The advantage of our model is especially evident when controlling signals in real-world cities, where road structures are more irregular and traffic flows are more dynamic. Specifically, \individualRL can hardly achieve satisfactory results because it independently optimizes the single intersection's policy; \oneModelNeighbor and \GCN do not work well for either \newYork or \hangzhou, as the agent treats the information from the upstream and downstream intersections with static importance according to the prior geographic knowledge rather than real-time traffic flows. 

2. The performance gap between the proposed \GAT and the conventional transportation method \maxPressure becomes larger as the evaluated data change from synthetic regular traffic (average gap $8.08\%$) to real-world dynamic traffic (average gap $19.89\%$). Such growing performance divergence conforms to the deficiency inherent in \maxPressure, that it is incapable of learning from the feedback of the environment.   


3. Our method outperforms the joint-action modelling method \nips. In order to achieve cooperation, \nips first builds up one model to decide the joint actions of two adjacent intersections and then uses centralized coordination over the global joint actions. It requires the centralized maximization over a combinatorially large joint action space and faces scalability issues. On the constrast, our method achieves cooperation through communication between decentralized agents, which has a smaller action space and shows superior performances.
\subsubsection{Convergence Comparison}
In Figure~\ref{fig:convergence}, we compare \GAT's performance (average travel time for vehicles evaluated at each episode) to the corresponding learning curves for the other five RL methods.
Evaluated in all the listed datasets, the performance of \GAT is better than any of the baselines by a large margin, both in \emph{jumpstart performance} (initial performance after the first episode), \emph{time to threshold} (learning time to achieve a pre-specified performance level), as well as in \emph{asymptotic performance} (final learned performance). Learning the attention on neighborhood does not slow down model convergence, but accelerates the speed of approaching the optimal policy instead. 

From Figure~\ref{fig:convergence_3x3}, we discover that model \individualRL starts with extremely huge travel time and approaches to the optimal performance after a long training time. Such disparity of convergence speed shown in Figure~\ref{fig:convergence} agrees with our previous space complexity analysis (in Section~\ref{sec:space_complexity}), that agents with shared models (\nips, \oneModelNeighbor, \oneModel, \GCN and \GAT) need to learn $O(m^2L)$ parameters while individual agents (\individualRL) have to update $O(m^2L\cdot N)$ parameters.

\subsection{Scalability Comparison}
\label{sec:exp:scala}
In this section, we investigate on whether \GAT is more scalable than other RL-based methods in the following aspects:

\subsubsection{Effectiveness.} As is shown in Table~\ref{tab:overall_performance} and the convergence curve in Figure~\ref{fig:convergence}, \GAT performs consistently better than other RL methods on networks with different scales, ranging from 9-intersection grid network to 196-intersection real-world network.
\subsubsection{Training time.} 
\begin{figure}[h]
\centering
\includegraphics[width=0.35\textwidth]{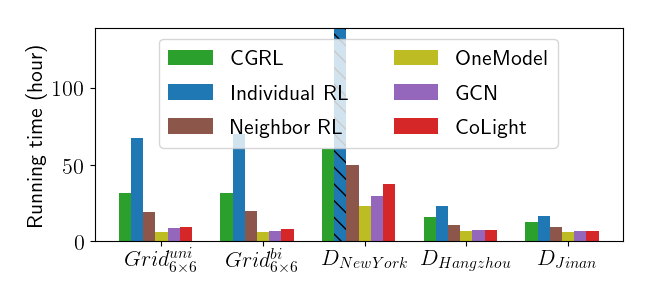}
\caption{The training time of different models for 100 episodes. \GAT is efficient across all the datasets. The bar for \individualRL on \newYork is shadowed as its running time is far beyond the acceptable time.}
\label{fig:running_time}
\vspace{-3mm}
\end{figure}

We compare \GAT's training time (total clock time for 100 episodes) to the corresponding running time for the other five RL methods on road networks with different scales. All the methods are evaluated individually on the server for fair comparison. As is shown in Figure~\ref{fig:running_time}, the training time for \GAT is comparable to that of \oneModel and \GCN, which is far more efficient than that of \nips, \individualRL and \oneModelNeighbor. This is consistent with the time complexity analysis (in Section~\ref{sec:time_complexity}), as most of the parallel computation assumptions are satisfied in our experiments.

Note that the average travel time for \individualRL (in Table~\ref{tab:overall_performance}) is missing and the bar of training time (in Figure~\ref{fig:running_time}) is estimated on \newYork setting. This is because \individualRL is non-scalable because all the separate $196$ agents cannot be trained and updated simultaneously due to processor and memory limitation. 

\subsection{Study of \GAT}
In this section, we investigate on how different components (i.e., neighborhood definition, number of neighbors, and number of attention heads) affect \GAT. 
\subsubsection{Impact of Neighborhood Definition.}
As mentioned in Section~\ref{sec:neighbor-scope}, the neighborhood scope of an intersection can be defined in different ways. And the results in Table~\ref{tab:overall_performance} show that \GAT (using geo-distance) achieves similar performance with \GATNode under synthetic data, but largely outperforms \GATNode under real-world traffic. The reason could be that under synthetic data, since the lane lengths of all intersections are the same, the top closest neighboring intersections set according to geo-distance is identical to that based on node distance. In the following parts of our experiments, we only compare \GATEuclidean with other methods.

\subsubsection{Impact of Neighbor Number.}
\begin{figure}[h!]
  \includegraphics[width=0.23\textwidth]{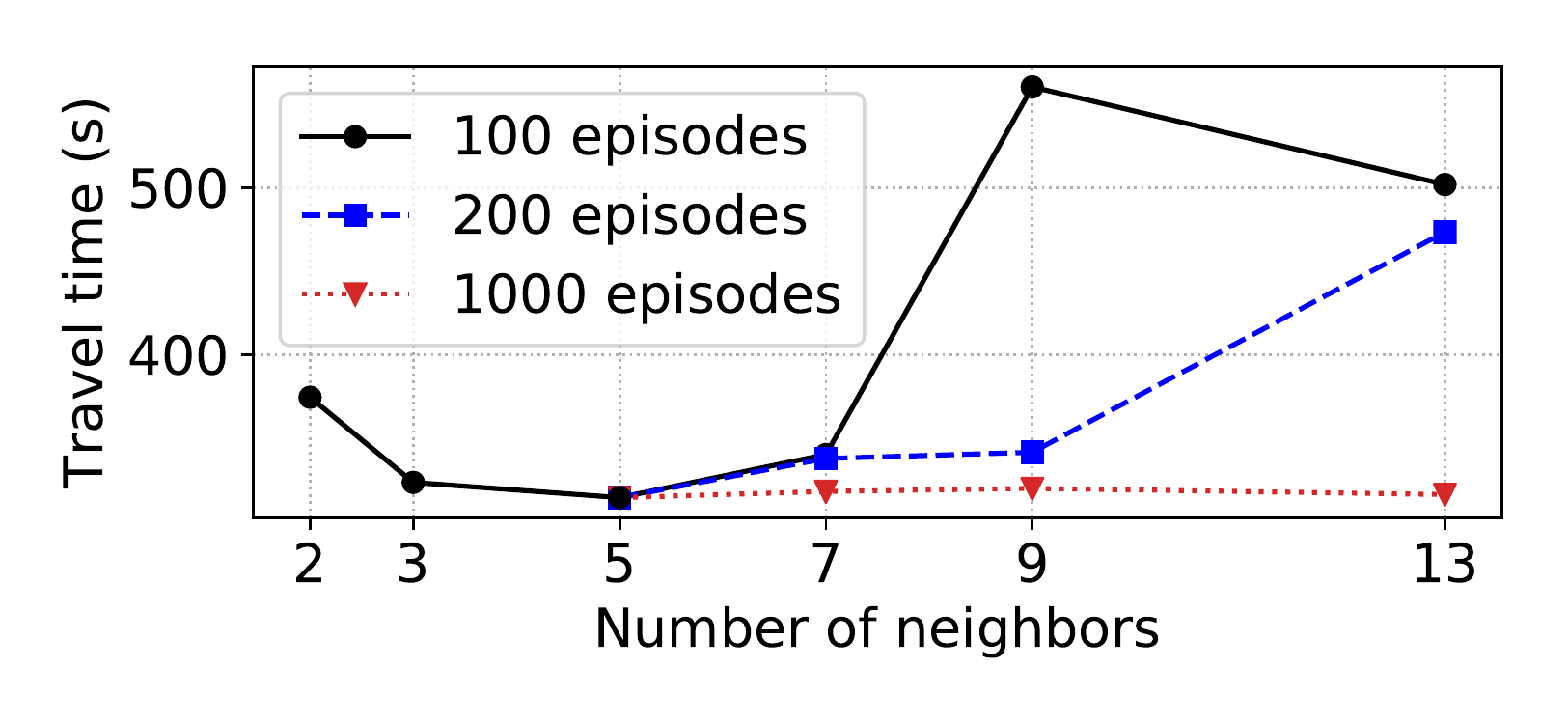}
  \includegraphics[width=0.23\textwidth]{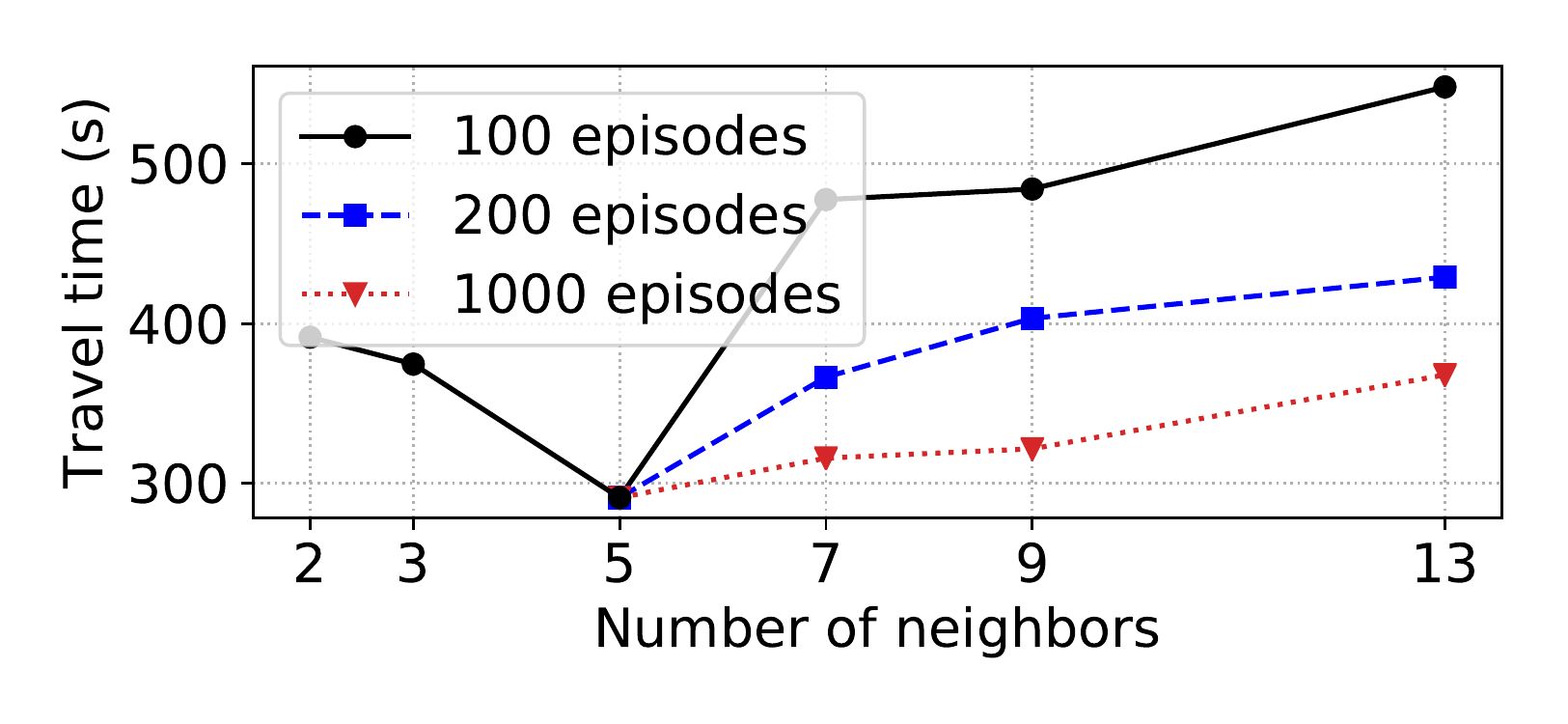}
\caption{Performance of \GAT with respect to different numbers of neighbors ($|\mathcal{N}_i|$) on dataset \hangzhou (left) and \jinan (right). More neighbors ($|\mathcal{N}_i|\leq 5$) for cooperation brings better performance, but too many neighbors ($|\mathcal{N}_i|>5$) requires more time (200 episodes or more) to learn.}\label{fig:para-neighbors}

\end{figure}
In Figure~\ref{fig:para-neighbors}, we show how the number of neighbors $|\mathcal{N}_i|$ influences the performance and also shed lights on how to set it. 
As is shown in Figure~\ref{fig:para-neighbors}, when the number of neighbors grows from 2 to 5, the performance of \GAT achieves the optimal. Further adding nearby intersections into the neighborhood scope $\mathcal{N}_i$, however, leads to the opposite trend. This is because including more neighbors in the neighborhood results in learning more relations. To determine signal control policy for each intersection, computing the attention scores only on four nearby intersections and itself seems adequate for cooperation with both time and performance guarantee.

\subsubsection{Impact of Attention Head Number.}
To evaluate the effectiveness of multi-head attention, we test different numbers of attention heads and find that moderate numbers of heads are beneficial to better control intersection signals. As shown in Table~\ref{tab:sense_heads}, drivers spend less time as the number of attention heads grows. However, the benefits of more types of attention disappear as $H$ exceeds $5$. Similar conclusions can be made on other datasets with details unshown due to space limitation.
\begin{table}[h]
\centering
\caption{Performance of \GAT with respect to different numbers of attention heads ($H$) on dataset \gridsix. More types of attention ($H\leq 5$) enhance model efficiency, while too many ($H>5$) could distract the learning and deteriorate the overall performance. }
\label{tab:sense_heads}
\begin{tabular}{cccccc}
\toprule
\#Heads& 1 & 3& 5& 7& 9 \\\midrule
Travel Time (s) &176.32 &172.47 &170.11 &174.54 &174.51\\\bottomrule
\end{tabular}
\end{table}

\section{Attention Study}
To analyze how well the neighborhood cooperation is implemented via the attention mechanism, we will study both the \emph{spatial} and \emph{temporal} distribution of attention learned by \GAT on both synthetic and real-world data. 
\subsection{Spatial Distribution}
The spatial distribution of attention scores indicates the importance of different neighbors to the target agent. In this section, we analyze the learned average attention that \GAT learns from the real-world data in each episode. We have the following observations: 

\begin{figure}[t]
  \centering
  \subfigure[Intersection A in New York]{\label{fig:attention_newyork_ny} \includegraphics[width=0.4\textwidth]{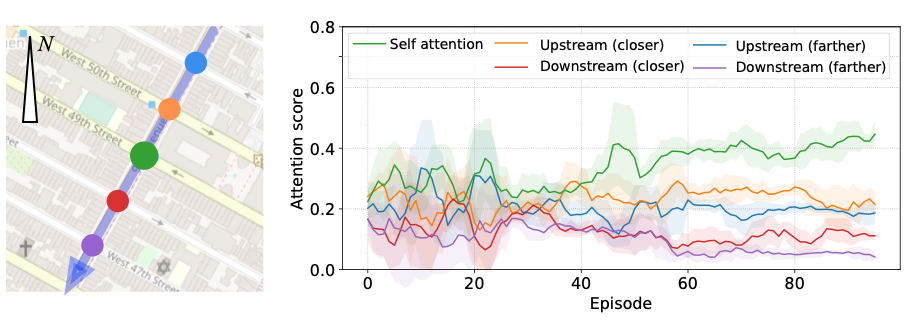}}
  \subfigure[Intersection B in Hangzhou]{\label{fig:attention_newyork_hz} \includegraphics[width=0.4\textwidth]{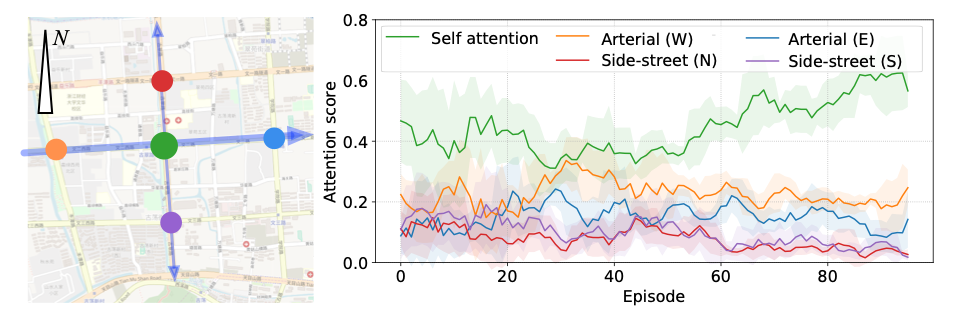}}
  \caption{Spatial difference of attention distribution learned by \GAT during training process in real-world traffic.
  Different colored lines in the right figures correspond with the colored dots in the left figures. 
  Up: For intersection A in \newYork, major concentration is allocated on upstream intersections and A itself. Down: For intersection B in \hangzhou, major concentration is allocated on arterial intersections and B itself.}
  \label{fig:attention_newyork}
\end{figure}

~\noindent\\$\bullet$\emph{Upstream vs. Downstream.} Figure~\ref{fig:attention_newyork_ny} shows an intersection (green dot) in New York, whose neighborhood includes four nearby intersections along the arterial. Traffic along the arterial is uni-directional (blue arrow). From the attention distribution learned by \GAT, we can see that while the majority of attention is allocated to itself, the upstream intersections (orange and blue lines) have larger scores than downstream intersections (red and purple lines).
~\noindent\\$\bullet$\emph{Arterial vs. Side Street.} Figure~\ref{fig:attention_newyork_hz} shows an intersection (green dot) in Hangzhou, whose neighborhood includes two intersections along the arterial and two intersections on the side street. In the left part of Figure~\ref{fig:attention_newyork_hz}, the arterial traffic is heavy and uni-directional (thick horizontal arrow), and the side-street traffic is light and bi-directional (thin vertical arrow). From the right part of Figure~\ref{fig:attention_newyork_hz}, we can see that the arterial intersections (orange and blue lines) have larger scores than side-street intersections (red and purple lines).
\subsection{Temporal Distribution}
\begin{figure}[h]
  \centering
  \subfigure[\gridthree road network]{\label{fig:temporal_network} \includegraphics[width=0.2\textwidth]{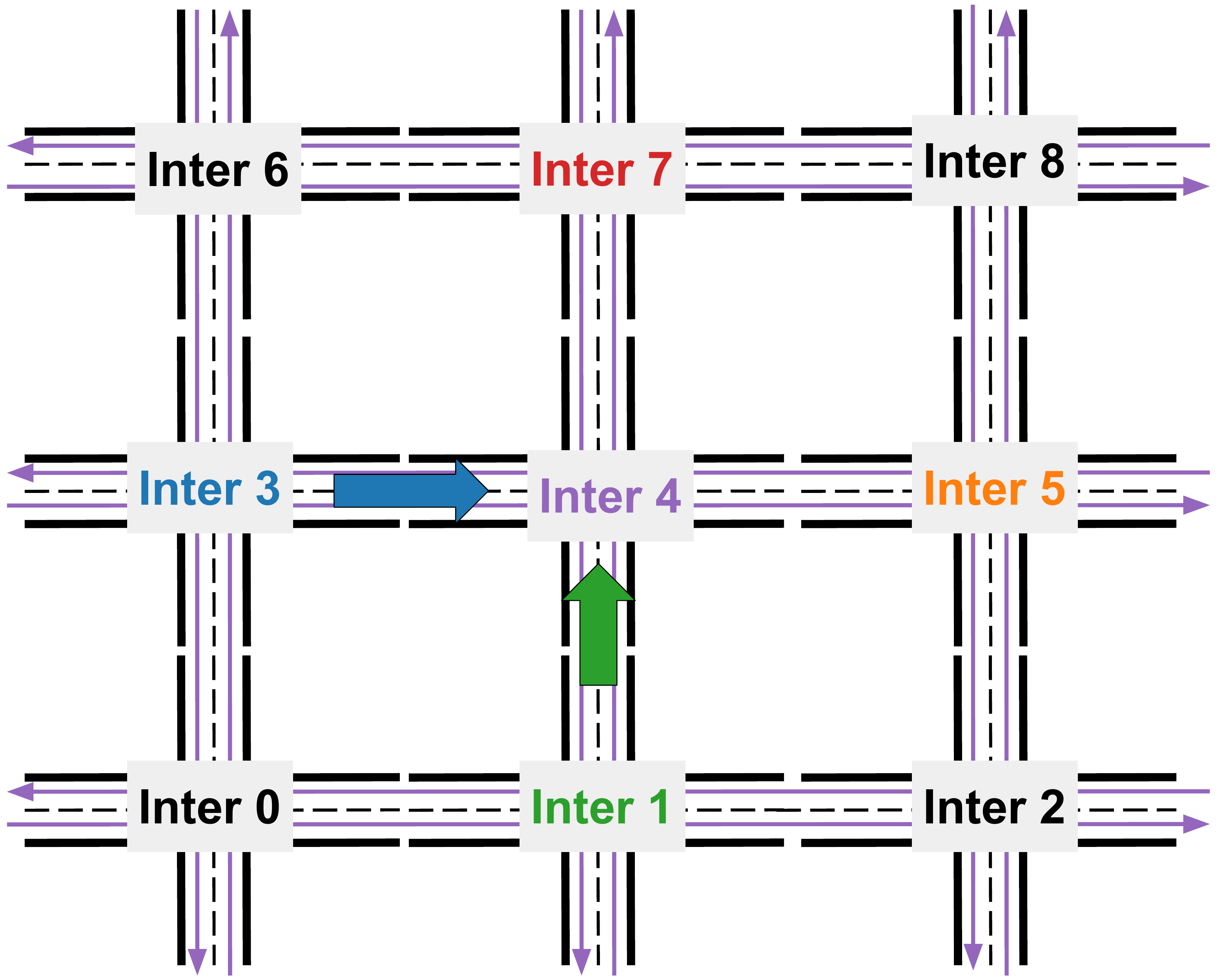}}
  \subfigure[Traffic flows and attention scores of neighbors to Inter\#4]{\label{fig:temporal_attention} \includegraphics[width=0.18\textwidth]{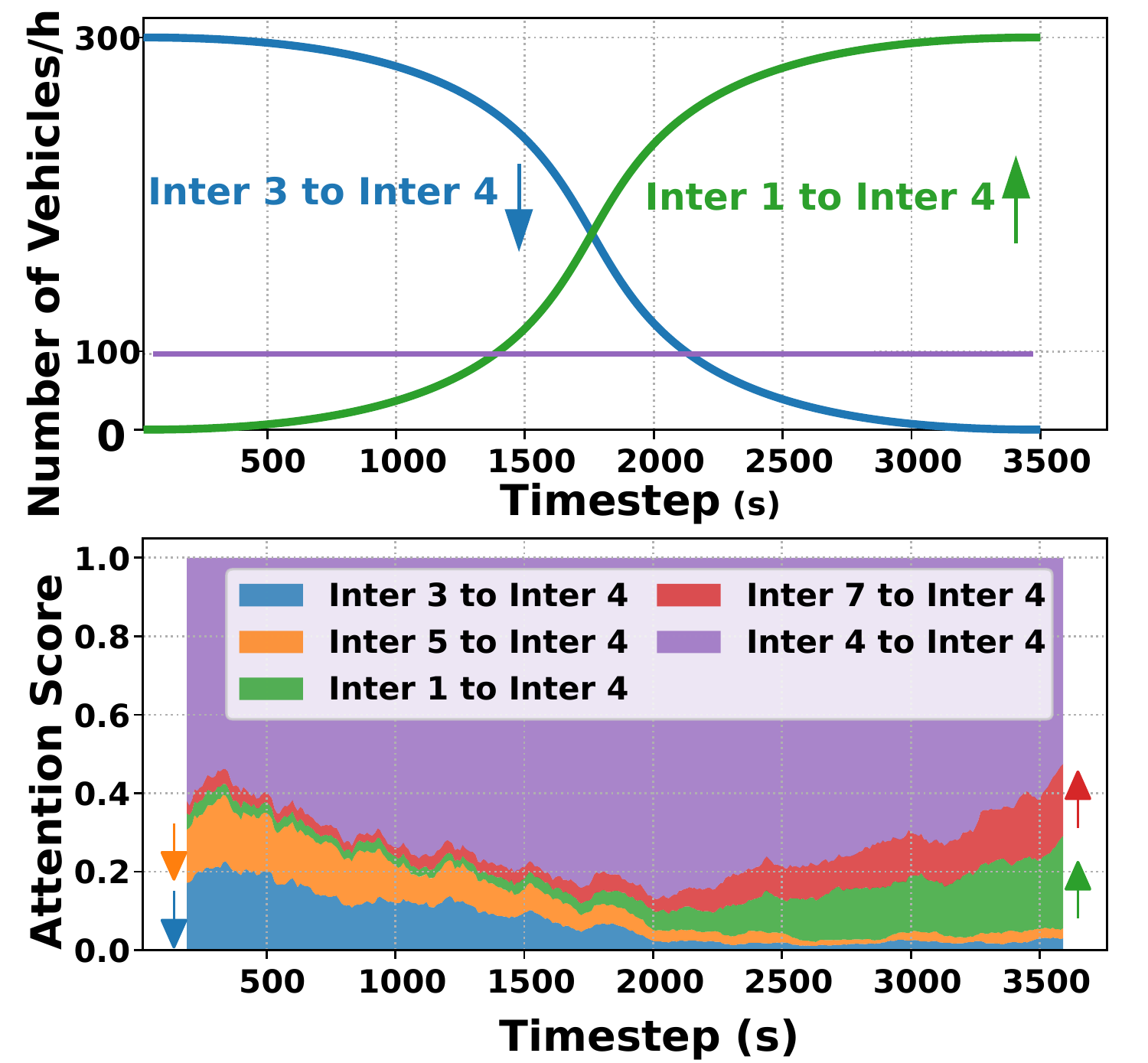}}
  \caption{Temporal distribution of attention score learned by \GAT corresponds with temporally changing traffic.}
  \label{fig:temporal_distribution}

\end{figure}

The temporal distribution of neighbors' attention scores throughout a certain time period indicates the temporal change of neighbors' influence. 
In this section, we analyze the learned attention of a converged \GAT model in a $3\times3$ network under a 3600-second traffic. As is shown in the upper part of Figure~\ref{fig:temporal_attention}, there is small basic traffic on every direction; at the same time, similar to the traffic change between morning peak hours and evening peak hours, the traffic of \emph{Inter \#4} experiences great changes, where the traffic from \emph{Inter \#3} to \emph{Inter \#4} drops (blue arrows in Figure~\ref{fig:temporal_network}) and the traffic from \emph{Inter \#1} to \emph{Inter \#4} increases (green arrows in Figure~\ref{fig:temporal_network}). 

The attention scores of the neighboring four intersections of \emph{Inter \#4} are shown in the lower part of Figure~\ref{fig:temporal_attention}. Firstly, we can see that the score from \emph{Inter \#4} to \emph{Inter \#4} always occupies the largest purple area, indicating that the traffic condition of \emph{Inter \#4} is of a great importance. Secondly, the attention scores of \emph{Inter \#1} and \emph{\#7} increase with more traffic on South$\leftrightarrow$North direction. Similarly, the attention scores of \emph{Inter \#3} and \emph{\#5} decrease as less traffic on West$\leftrightarrow$East direction. The temporal change of attention scores of \emph{Inter \#4}'s four neighbors matches exactly with the change of traffic flow from the four directions, which demonstrates that the proposed \GAT model is capable to capture the key neighborhood information over time, i.e., the temporal attention.
\section{Conclusion}

In this paper, we propose a well-designed reinforcement learning approach to solve the network-level traffic signal control problem. Specifically, our method learns the dynamic communication between agents and constructs an index-free model by leveraging the graph attention network. We conduct extensive experiments using synthetic and real-world data and demonstrate the superior performance of our proposed method over state-of-the-art methods. In addition, we show in-depth case studies and observations to understand how the attention mechanism helps cooperation.

We would like to point out several important future directions to make the method more applicable to the real world. First, the neighborhood scope can be determined in a more flexible way. The traffic flow information between intersections can be utilized to determine the neighborhood. Second, the raw data for observation only include the phase and the number of vehicles on each lane. More exterior data like the road and weather condition might help to boost model performance.

\begin{acks}
The work was supported in part by NSF awards \#1652525 and \#1618448. The views and conclusions contained in this paper are
those of the authors and should not be interpreted as representing
any funding agencies.
\end{acks}

\bibliographystyle{ACM-Reference-Format}
\bibliography{proc}


\end{document}